\documentclass[11pt]{article}
\usepackage[top=1.25in, bottom=1.25in, left=1.25in, right=1.25in]{geometry}
\usepackage{amssymb}
\usepackage{latexsym}
\usepackage{amsmath}
\usepackage{amsthm}
\usepackage{mathtools}
\usepackage{tikz-cd}
\usepackage{tikz}
\usepackage{stmaryrd}
\usepackage{enumitem} 
\usepackage{hyperref}
\usepackage{graphicx}
\usepackage{caption}
\usepackage{multicol}
\usepackage{subcaption}
\usepackage[title]{appendix}
\usepackage{mathrsfs}
\usepackage{bbm}
\usepackage{bm}
\usepackage{color,soul}
\usepackage{stackengine, scalerel}
\usepackage[font=footnotesize]{caption}

\graphicspath{ {./images/} }
\hypersetup{
    colorlinks,
    citecolor=blue,
    filecolor=red,
    linkcolor=blue,
    urlcolor=blue
}

\theoremstyle{definition}

\theoremstyle{definition}

\title{Persistent topological Laplacian analysis of SARS-CoV-2 variants}
\author{Xiaoqi Wei$^{1}\footnote{
 E-mail: weixiaoq@msu.edu} $, 
 Jiahui Chen$^{1}\footnote{E-mail: chenj159@msu.edu}$,
 and Guo-Wei Wei$^{1,2,3}\footnote{Corresponding author.
 E-mail: weig@msu.edu}$ \\
$^1$ Department of Mathematics, \\
Michigan State University, MI 48824, USA.\\
$^2$ Department of Electrical and Computer Engineering,\\
Michigan State University, MI 48824, USA. \\
$^3$ Department of Biochemistry and Molecular Biology,\\
Michigan State University, MI 48824, USA. \\
}
\date{}

\begin{document}
\maketitle
\begin{abstract}
Topological data analysis (TDA) is an emerging field in mathematics and data science. 
Its central technique, persistent homology, has had tremendous success in many science and engineering disciplines. 
However, persistent homology has limitations, including its inability to handle heterogeneous information, such as multiple types of geometric objects;  being qualitative rather than quantitative, e.g., counting a 5-member ring the same as a 6-member ring, and a failure to describe non-topological changes, such as homotopic changes in protein-protein binding.  
Persistent topological Laplacians (PTLs), such as persistent Laplacian and persistent sheaf Laplacian, were proposed to overcome the limitations of persistent homology.   
In this work, we examine the modeling and analysis power of PTLs in the study of the protein structures of the severe acute respiratory syndrome coronavirus 2 (SARS-CoV-2) spike receptor binding domain (RBD). 
First, we employ PTLs to study how the RBD mutation-induced structural changes of RBD-angiotensin-converting enzyme 2 (ACE2) binding complexes are captured in the changes of spectra of the PTLs among SARS-CoV-2 variants. 
Additionally, we use PTLs to analyze the binding of RBD and ACE2-induced structural changes of various SARS-CoV-2 variants. 
Finally, we explore the impacts of computationally generated RBD structures on a topological deep learning paradigm 
and predictions of deep mutational scanning datasets for the SARS-CoV-2 Omicron BA.2 variant.   
Our results indicate that PTLs have advantages over persistent homology in analyzing protein structural changes and provide a  powerful new TDA tool for data science. 
\end{abstract}


\tableofcontents

\section{Introduction}

Severe acute respiratory syndrome coronavirus 2 (SARS-CoV-2) is 
the cause of the ongoing global coronavirus disease 2019 (COVID-19) pandemic. 
Its evolution and future direction are of major concern. 
It was well established that the emergence of SARS-CoV-2 new variants is dictated by mutation-induced infectivity strengthening \cite{chen2020mutations} and antibody resistance (or vaccine breakthrough) \cite{wang2021mechanisms}, two molecular mechanisms that determined the natural selection at the population scale. 
More specifically, the binding of the viral spike protein, particularly 
the receptor-binding domain (RBD), to the human receptor angiotensin-converting enzyme 2 (ACE2) facilitates the entry of the virus into host cells ~\cite{hoffmann2020sars,walls2020structure}. 
In early 2020, it was hypothesized that natural selection favors those SARS-CoV-2 RBD mutations that strengthen the RBD-ACE2 binding, which leads to higher viral infectivity ~\cite{chen2020mutations}. 
The hypothesis was initially supported by the frequency analysis of 89 single RBD mutations found from the genotyping of 15,140 complete SARS-CoV-2 genome samples \cite{chen2020mutations} and later confirmed beyond doubt by the evolution pattern of 651 RBD mutations found from the genotyping of 506,768  SARS-CoV-2 genomes extracted from COVID-19 patients up to early 2021~\cite{wang2021vaccine}.

The vaccine breakthrough mechanism was not discovered until vaccines became widely available in industrialized countries in the summer of 2021. 
It was found that an RBD mutation that weakens the viral infectivity had an unusually high observed frequency in  2,298,349 complete SARS-CoV-2 genomes isolated from
patients. 
This abnormal statistics was found to strongly correlate with the vaccination rates in a few industrialized countries, including Denmark, the United Kingdom, France, Bulgaria, the United States, etc \cite{wang2021mechanisms}.    
To understand this correlation, the mutational impact of a set of 130 antibodies extracted from Covid patients that targets the RBD was studied.  
It was found that the abnormal mutation on the RBD has a very strong ability to disrupt the binding of most antibody-RBD complexes, 
which gives rise to antibody resistance (or vaccine breakthrough) at the population scale  \cite{wang2021mechanisms}.   

As discussed above, the reveal of the natural selection mechanisms of SARS-CoV-2 evolution is a typical example of a data-driven discovery that cannot be achieved by individual experimental laboratories. 
In fact, the discovery utilized results from tens of thousands of experimental laboratories around the world  \cite{chen2020mutations,wang2021mechanisms}. 
Machine learning, including deep learning, and also data-driven approach, played an essential role in the discovery. 
Deep learning methods can offer some of the most accurate predictions of biomolecular properties, 
including the binding affinity of protein-protein interactions (PPIs). 
This approach becomes particularly advantageous and outperforms other methods when good-quality experimental data are available. 
However, structure-based machine learning, including deep learning methods encounter difficulties in PPI predictions due to their intricate structural complexity and high dimensionality.

Advanced mathematics, such as topological data analysis (TDA), can provide an effective abstraction of PPIs \cite{wang2020topology}. 
TDA is an emerging mathematical field that utilizes algebraic topology approaches to analyze data. 
Its main tool is persistent homology~\cite{carlsson2009topology,edelsbrunner2000topological,frosini1992measuring,kovacev2016using,liu2023neighborhood,townsend2020representation,KLXia:2014c,zomorodian2005computing}, 
which integrates classical homology and filtration to create a multiscale analysis of data, resulting in a family of topological invariants. Through analyzing the signature and change of topological invariants during filtration, one can infer the shape of data \cite{carlsson2009topology}. 
However, persistent homology has limitations. Firstly, it is insensitive to homotopic shape evolution that does not involve any topological change. Secondly, roughy speaking, it cannot distinguish between a five-member ring and a six-member ring. Thirdly, it is incapable of differentiating different types of atoms, unable to describe directed relations, and indifferent to structured data such as functional groups.
To overcome the first two limitations, persistent spectral graph  \cite{wang2020persistent}, also known as  persistent Laplacian \cite{memoli2022persistent,wang2021hermes}, was proposed. This method not only returns the full set of topological invariants as persistent homology does but also captures additional homotopic shape evolution of data and is more quantitative in its non-harmonic spectra.
In addition to mathematical analysis~\cite{memoli2022persistent}, computational algorithms, such as HERMES software package \cite{wang2021hermes} and homotopy continuation \cite{wei2021homotopy}, were developed to facilitate topological deep learning, an emerging paradigm first introduced in 2017 \cite{cang2018representability,cang2017topologynet} for biomolecular studies, i.e.,  persistent Laplacian-assisted protein-ligand binding~\cite{meng2021persistent} and protein-protein binding~\cite{chen2022toplapnet,wee2022persistent}. 
Neither persistent homology nor persistent Laplacian is sensitive to heterogeneous information in data. The element-specific persistent homology was designed to alleviate this difficulty. This approach has had tremendous success in deciphering biomolecules \cite{cang2017topologynet, cang2018representability} and 
in worldwide computer-aided drug design competitions \cite{nguyen2019mathematical}.
Inspired by this success, various new TDA methods have been proposed \cite{ameneyro2022quantum,grbic2022aspects, liu2022biomolecular,liu2022hom}. 
Recently, a more elegant theory, persistent sheaf Laplacian, was proposed to embed heterogeneous information, such as geometry and partial charges, in topological analysis \cite{wei2021persistent}, 
utilizing the theory of cellular sheaves \cite{hansen2019toward,JustinCurry}. Both persistent Laplacian and persistent sheaf Laplacian belong to a class of persistent topological Laplacians (PTLs) \cite{wei2023topological}.  
PTLs are a family of multiscale topological spectral methods, including continuous (evolutionary) Hodge Laplacians defined on manifolds \cite{chen2021evolutionary} and all other discrete multiscale topological Laplacians, namely, persistent sheaf Laplacians \cite{wei2021persistent}, persistent spectral graphs \cite{wang2020persistent}, persistent path  Laplacians \cite{wang2023persistent},  persistent topological hypergraph Laplacians \cite{chen2023persistent},   persistent  hyperdigraph Laplacians \cite{chen2023persistent}, etc. 
Among them, persistent path  Laplacians were designed to describe directed graphs (digraphs) and directed networks, while persistent topological hypergraph Laplacians and persistent  hyperdigraph Laplacians can further deal with structured data.  
These new TDA methods can generate efficient mathematical representations of macromolecules either being used to model molecular structures or being used jointly with machine learning models for predicting various properties of molecules \cite{chen2022toplapnet}.
In the past three years, TDA approaches have been applied to   SARS-CoV-2 related databases to predict PPI binding free energy (BFE) changes of RBD-ACE2 and RBD-antibody complexes induced by RBD mutations~\cite{chen2021prediction,chen2021revealing}. Particularly, the non-harmonic spectra of PTLs can further unveil the homotopic geometric deformation induced by RBD mutations.
 
Although sequence-based approaches offer good predictions of mutational impacts on proteins, structure-based methods outperform other approaches \cite{qiu2022persistent}. In machine-learning-assisted directed evolution and protein engineering and machine-learning-based PPI and protein folding stability predictions, mutant structures are typically not available and are conventionally created by computational means for the machine learning predictions \cite{cang2017topologynet, cang2018representability, chen2021prediction, chen2021revealing,liu2022hom}, which is a source of errors. It is interesting and important to quantify such errors. Fortunately, since SARS-COV-2 variants are some of the most studied subjects, some of their three-dimensional (3D) structures are available in the literature, which offers an opportunity for in-depth analysis and comparison.   
  
Our objectives for this work are three-fold.  
We are interested in both the structural changes of the wild type RBD induced by mutations and the structural changes of the wild type RBD or  mutant RBDs induced by their binding to ACE2. To quantify structural changes we first perform alignment of structures and   calculate the distances between corresponding atoms (e.g., C$_\alpha$). Then, we compute PTLs of different structures to further characterize their structural changes.
Finally, we study how the difference between experimentally determined mutant structures and computationally generated mutant structures affects PTL-based machine learning and topological deep learning  predictions of PPIs.
  This is important because we want to understand  machine learning models'  stability with respect to structural perturbations and approximations. 
To this end, we utilize the 3D structures of SARS-CoV-2 RBD-ACE2 complexes of the wild type and mutants such as Alpha, Beta, Gamma, and Omicron BA1 and BA2. 
We also employed the 3D spike protein structures of the wild type and mutants such as Alpha, Beta, and Omicron BA.1 and BA.2. 
Persistent Laplacian and persistent sheaf Laplacian are tested in our studies.
To quantitatively analyze the influence of computationally generated structures on machine learning models, we used two topological machine learning models, namely TopLapGBT and TopLapNet \cite{chen2022toplapnet}  and a deep mutational scanning (DMS) dataset based on Omicron BA.2 \cite{starr2022deep}. 
We found that for this dataset, the effects introduced by computationally generated structure on TopLapGBT are not significant. However, they may slightly reduce the accuracy of TopLapNet.  

\section{Results}

\begin{figure}[htbp]
    \centering
    \includegraphics[width=1.0\linewidth]{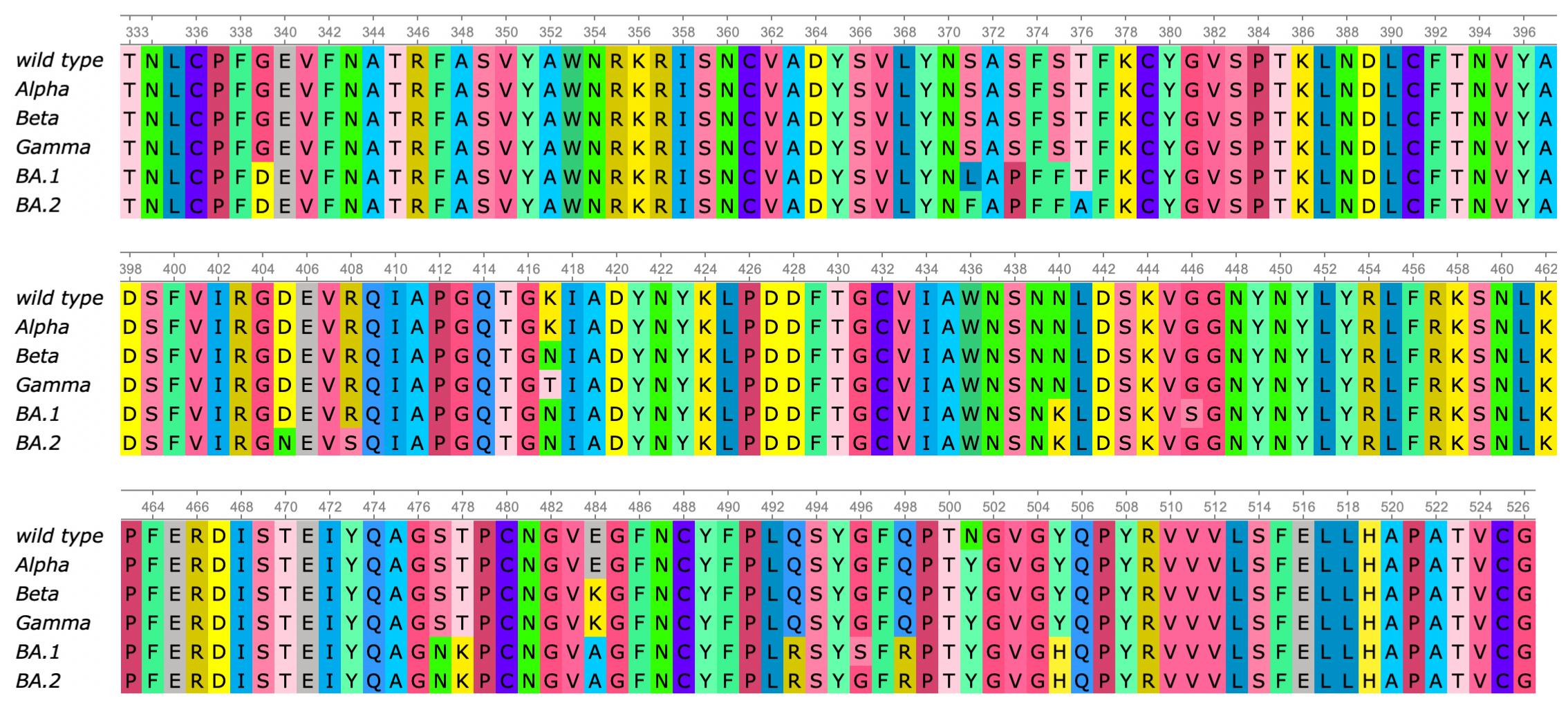}
    \caption{ Sequence alignment of RBDs of the wild type, Alpha, Beta, Gamma, BA.1, and BA.2.
    Alpha has one RBD mutation N501Y. Beta has three RBD mutations K417N, E484K, and N501Y.
    Gamma has three RBD mutations K417T, E484K, and N501Y. 
    BA.1 has 15 RBD mutations G339D, S371L, S373P, S375F, K417N, N440K, G446S, S477N, T478K, E484A, 
    Q493R, G496S, Q498R, N501Y, and Y505H. 
    BA.2 has 16 RBD mutations G339D, S371F, S373P, S375F, T376A, D405N, R408S, K417N, N440K, S477N, T478K,
    E484A, Q493R, Q498R, N501Y, and Y505H.}
    \label{seqalignment}
\end{figure}

\subsection{PTL analysis of RBD structural changes induced by mutations}

\begin{figure}[htbp]
    \centering
    \includegraphics[width=0.8\linewidth]{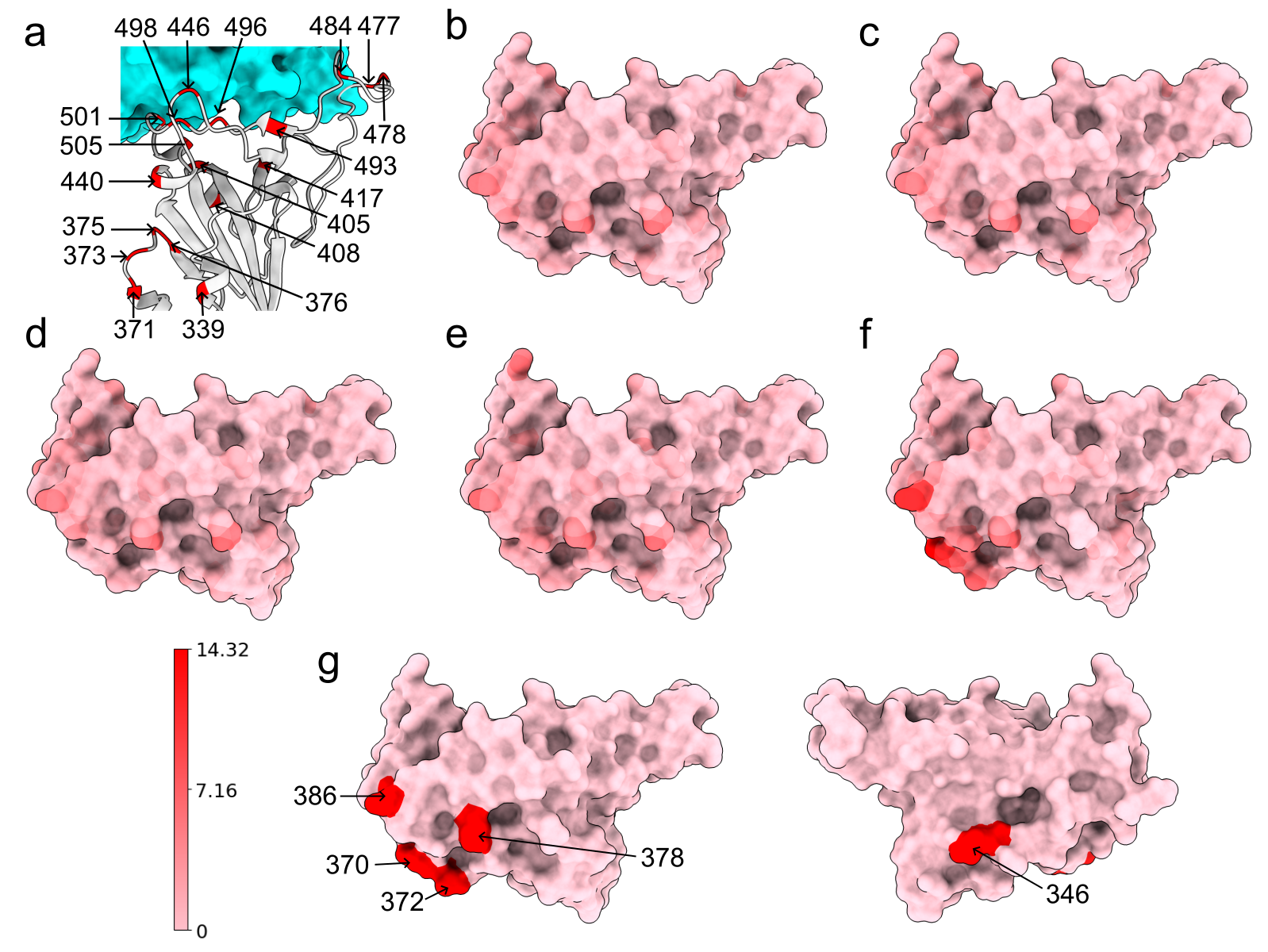}
    \caption{(a) Wild type RBD-ACE2 complex. 
    The RBD is colored by light grey and mutated residues in Alpha, Beta, Gamma, BA.1 and BA.2 are marked. 
    (b, c, d, e, f) Atoms of the wild type RBD are colored by their distances to corresponding atoms in a mutant RBD.
    Subfigures (a), (b), (c), (b), and (f) corresponds to the Alpha, Beta, Gamma, BA.1, and BA.2 variants, respectively.
    Pink and red corresponds to 0\AA\ and 14.32\AA\ respectively. 
    For each mutant we record the residues that have at least one atom whose distance to the corresponding atom in the wild type RBD is larger than 7.16\AA.
    In Alpha, Beta, and Gamma, such residue is R346. 
    In BA.1, such residue is K386. 
    In BA.2, such residues are N370, A372, K378, and K386. 
   These residues are marked in (g). 
    (Plots generated by ChimeraX \cite{pettersen2021ucsf}.)}
    \label{alignment}
\end{figure}

\begin{figure}[htbp]
    \centering
    \includegraphics[width=0.65\linewidth]{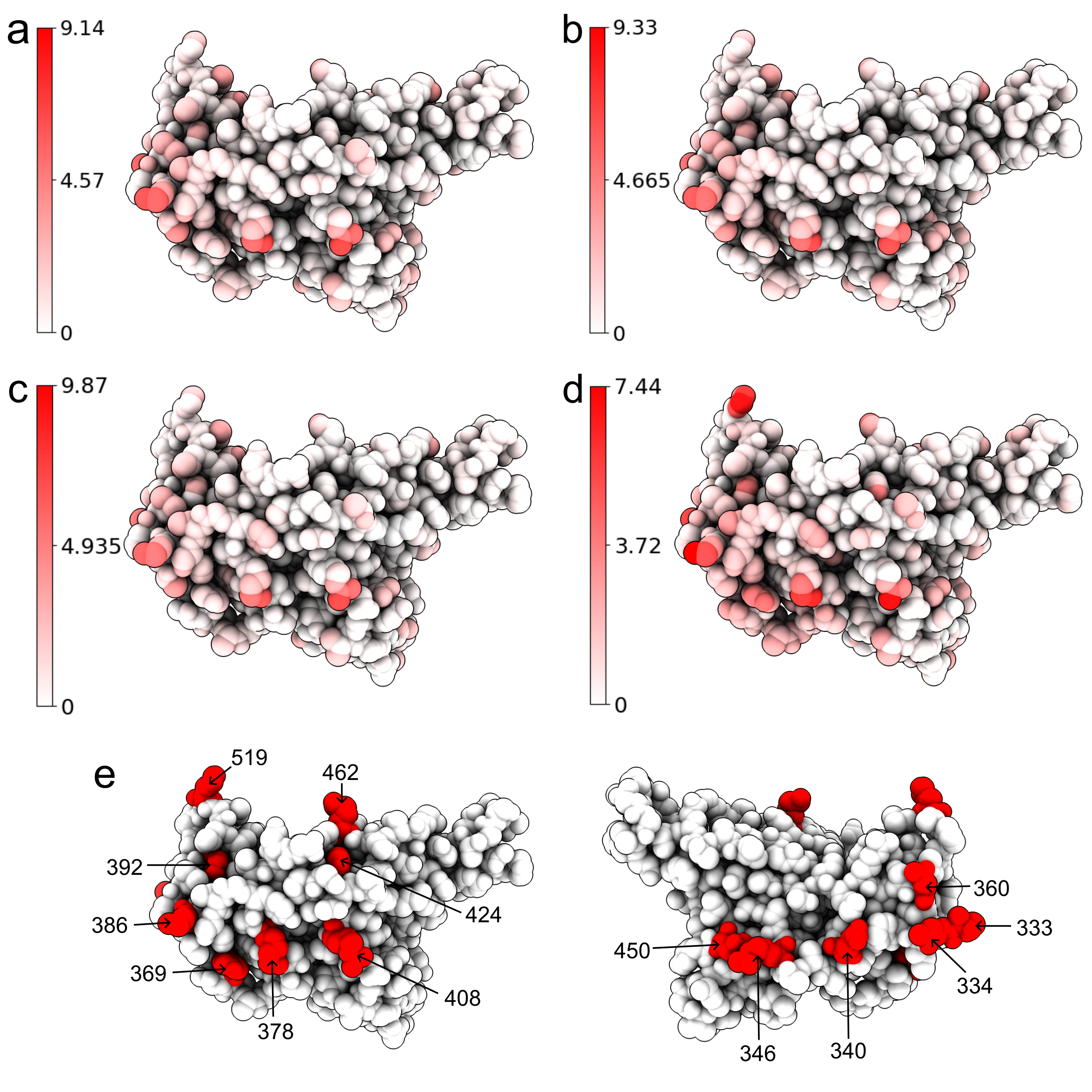}
    \caption{Atoms of the wild type RBD are colored by their distances to corresponding atoms in a mutant RBD.
    Subfigures (a), (b), (c), and (b) corresponds to the Alpha, Beta, Gamma, and BA.1, variants, respectively.
    Each alignment has its own color range. 
    For each mutant, we record the residues that have at least one atom whose distance to the corresponding atom in the wild type RBD is more than half of the maximal distance (4.57\AA, 4.67\AA, 4.94\AA, and 3.72\AA) between corresponding atoms.
    In Alpha, such residues are T333, R346, K378, K386, R408, and N450.
    In Beta and Gamma, such residues are T333, R346, K378, K386, and R408.
    In BA.1, such residues are T333, N334, E340, R346, N360, D364, Y369, K378, K386, F392, R408, K424, N450, K462, and H519.
  These residues are marked in (e). 
    (Plots generated by ChimeraX \cite{pettersen2021ucsf}.)}
    
    \label{alignment_local}
\end{figure}

To understand the structural differences of RBD between the wild type and mutants in RBD-ACE2 complex, we align the RBDs of SARS-CoV-2 variants Alpha (PDB ID: 8DLK\cite{mannar2022sars}), Beta (PDB ID: 8DLN\cite{mannar2022sars}), Gamma (PDB ID: 8DLQ\cite{mannar2022sars}), BA.1 (PDB ID: 7T9L\cite{mannar2022omicron}), and BA.2 (PDB ID: 7XB0\cite{li2022structural}) along with the wild type RBD (PDB ID: 6M0J\cite{lan2020structure})  in Figures \ref{alignment} and \ref{alignment_local}.
For Alpha, Beta, Gamma, BA.1, and BA.2, the maximal distances between corresponding atoms of mutant RBDs and the wild-type RBD are 9.14\AA, 9.33\AA, 9.87\AA, 7.44\AA, and 14.32\AA\ respectively. 
For each mutant, the residues are recorded if they have at least one atom whose distance to the corresponding atom in wild-type RBD is more than 7.16\AA, which is half of the maximal distance,  14.32\AA.   
For variants Alpha, Beta, and Gamma, such a residue is R346, while in BA.1 such residue is K386. 
BA.2 has most such residues, which are N370, A372, K378, and K386, containing atoms deviating from the wild type.
However, these residues are not in the receptor-binding motif (RBM, residues 438-506) that interacts directly with ACE2.

Alternatively, for Alpha, Beta, Gamma, and BA.1 variants, we also change the threshold from 7.16\AA\ to the half of maximal distance (4.57\AA, 4.67\AA, 4.94\AA, and 3.72\AA, respectively).
Then in the Alpha variant, such residues are T333, R346, K378, K386, R408, and N450.
In Beta and Gamma variants, such residues are T333, R346, K378, K386, and R408.
In BA.1 such residues are T333, N334, E340, R346, N360, D364, Y369, K378, K386, F392, R408, K424, N450, K462, and H519. 
Also most large C$_\alpha$ structural changes occur at the coil regions of the RBD.
For the BA.2 variant, the half of maximal distance is 7.16\AA\, and we have recorded such residues that have at least one atom whose distance to the corresponding atom in the wild-type RBD is more than 7.16\AA.

\begin{figure}[htbp]
  \centering
  \includegraphics[width=0.8\linewidth]{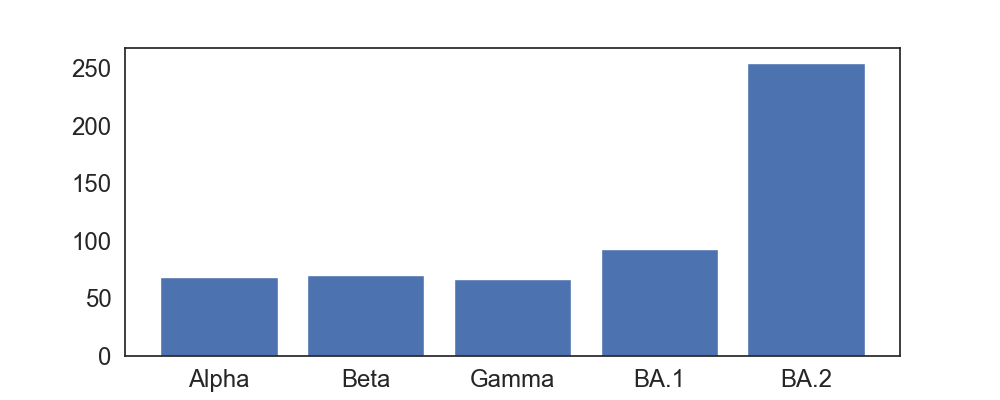}
  \caption{The total structural changes of RBD between the wild type and mutants in RBD-ACE2 complex.
  Given an alignment of a mutant RBD to the wild type RBD, 
  the total structural changes is defined to be the sum of squares of distances between corresponding 
  C$_\alpha$ atoms in RBD.
  }
  \label{total_changes}
\end{figure}

To quantify the total structural differences between the wild type and mutants, 
we calculate the sum of squares of distances between corresponding C$_\alpha$ atoms. The results of Alpha, Beta, Gamma, BA.1, and BA.2 are 69 \AA$^2$, 70 \AA$^2$, 67\AA$^2$, 93\AA$^2$, and 255\AA$^2$, respectively as shown in Figure \ref{total_changes}. The large  values for BA.1 and BA.2 are consistent with fact that BA.1 and BA.2 are strongly antibody disruptive  \cite{chen2022omicron,chen2022omicron2}. The large structural changes induced by BA.2 mutations create significant mismatch between antibodies and antigens, making BA.2 one of the most antibody resistant variants \cite{chen2022omicron2}. 
Arguably, the amount of mutation-induced structural changes in  RBD-ACE2 complexes also strongly correlates with viral infectivity changes.

\begin{figure}[htbp]
    \centering
        \includegraphics[width=0.9\linewidth]{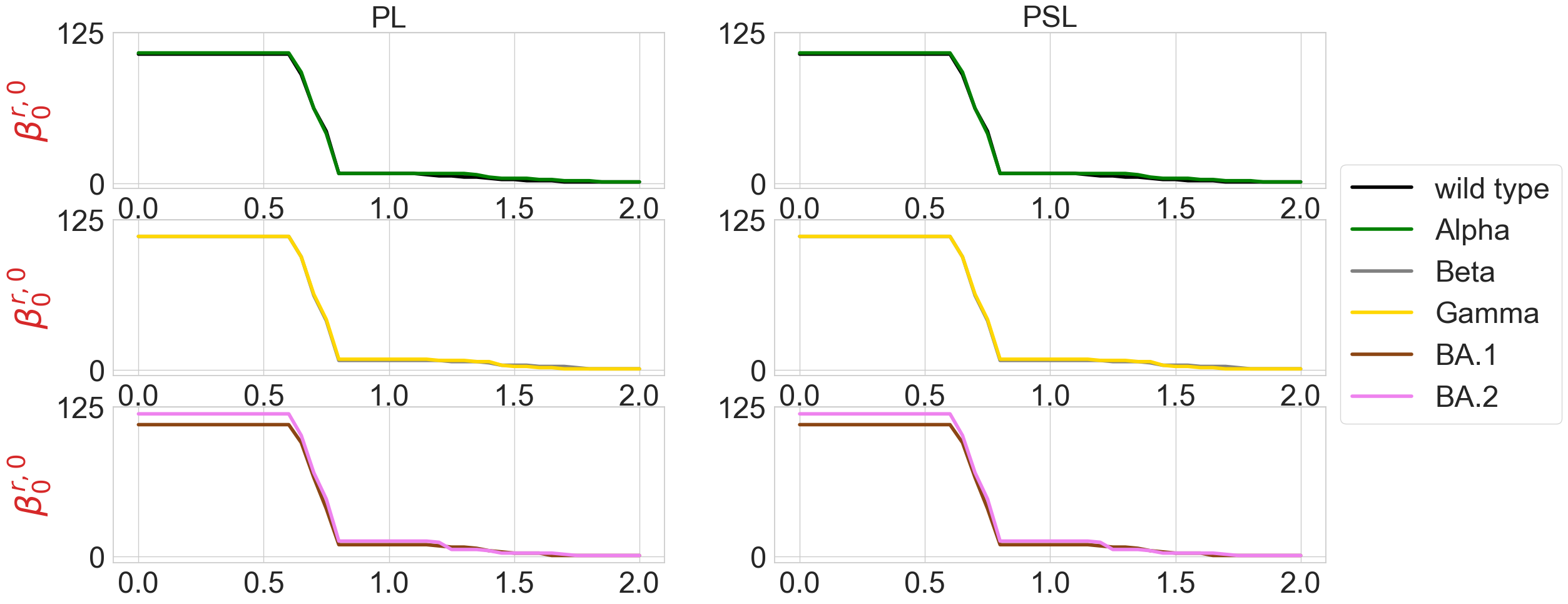}  
    \caption{Illustration of persistent (sheaf) Betti numbers of element nonspecific persistent  Laplacian (PL) and persistent sheaf   Laplacian (PSL) 
    of the residue 501 mutation site at different filtration values, i.e., radii (unit: \AA). 
		The wild type (PDB ID: 6M0J) and Alpha (PDB ID: 8DLK) are given in the first row. 
		The Beta (PDB ID: 8DLN) and Gamma (PDB ID: 8DLQ) are given in the second row. 
		BA.1 (PDB ID: 7T9L) and BA.2 (PDB ID: 7XB0) are given the third row.
    }
    \label{mutation_nes_beta}
\end{figure}

\begin{figure}[htbp]
    \centering
        \includegraphics[width=0.9\linewidth]{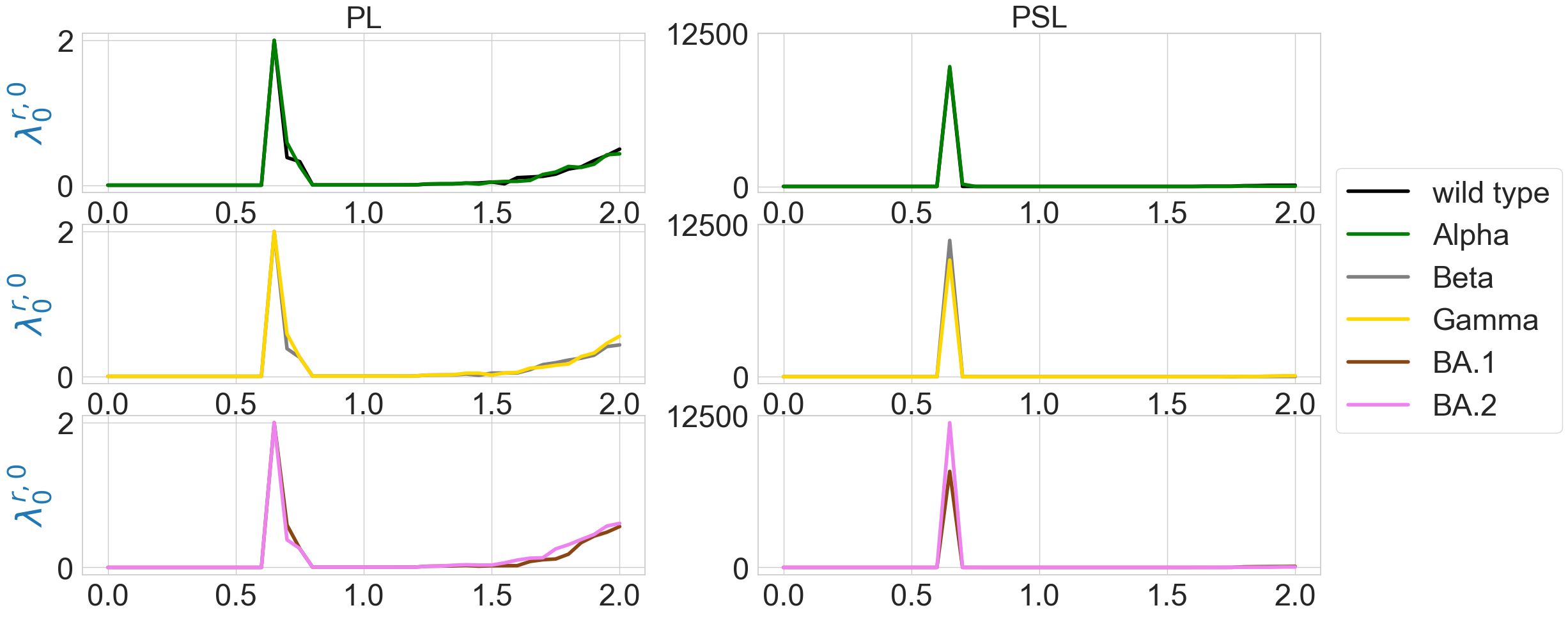}
    \caption{Illustration of the first nonzero eigenvalues  of element nonspecific persistent   Laplacian (PL) and persistent sheaf   Laplacian (PSL) 
    of the residue 501 mutation site at different filtration values, i.e., radii (unit: \AA). 
		The wild type (PDB ID: 6M0J) and Alpha (PDB ID: 8DLK) are given in the first row. 
		The Beta (PDB ID: 8DLN) and Gamma (PDB ID: 8DLQ) are given in the second row. 
		BA.1 (PDB ID: 7T9L) and BA.2 (PDB ID: 7XB0) are given the third row.
		}
    \label{mutation_nes_lambda}
\end{figure}

\begin{figure}[htbp]
    \centering
    \includegraphics[width=0.8\linewidth]{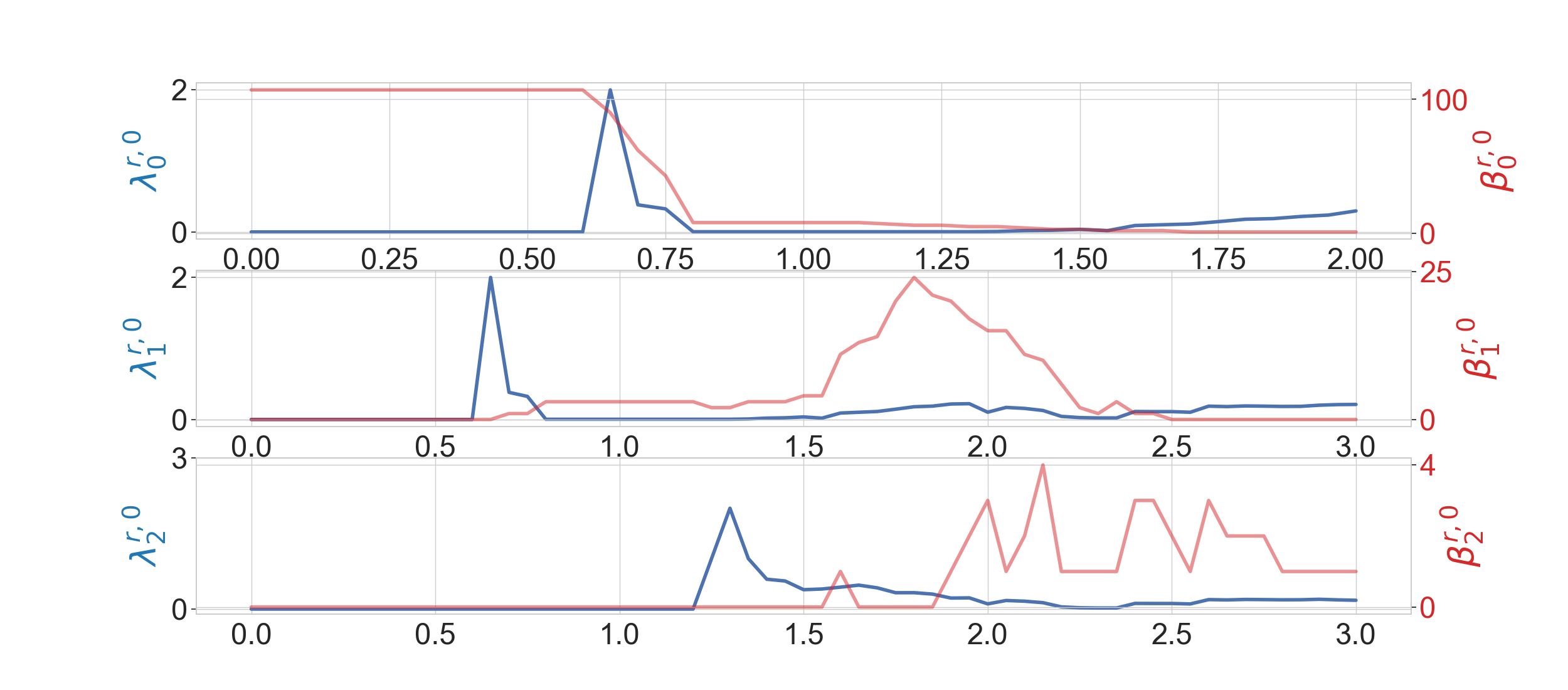}
    \caption{Illustration of persistent Betti numbers (red line) and the first nonzero eigenvalues (blue line) of element nonspecific persistent Laplacians 
    of the wild type N501 mutation site at different filtration values, i.e., radii (unit: \AA). Alpha filtration is used.
    The graphs from top to bottom represent the results of dimension-0, dimension-1, and dimension-2 Laplacians.}
    \label{miltidimen}
\end{figure}
We are also interested in the topological characterization of the mutation-induced conformational changes. To this end, we employ persistent Laplacian (PL) and persistent sheaf Laplacian (PSL) to examine the local RBD structural changes induced by the mutation N501Y (a common mutation that exists in Alpha, Beta, Gamma, BA.1, and BA.2).
For the wild type and mutants, the residue 501 mutation site
is defined as the set of neighborhood heavy atoms (C, N, and O) in RBD such that the distance of any atom in the set to the residue 501 C$_\alpha$ is smaller than 10\AA. 
We calculate persistent Laplacians and persistent sheaf Laplacians for mutation sites of the wild type and variants and compare the persistent (sheaf) Betti number and the smallest nonzero eigenvalues of spectra at different filtration values. 
Persistent Laplacians and persistent sheaf Laplacians can be calculated as either element non-specifically or element specifically (i.e., considering carbon, nitrogen, and oxygen atoms separately).
We first employ the element non-specific approach and compare the results of the wild type and variants. 
The results of persistent Laplacian and persistent sheaf Laplacian are shown in Figures \ref{mutation_nes_beta} and \ref{mutation_nes_lambda}.
The $x$ axis represents the filtration values of Rips filtration, such that at a filtration value $r$ the Rips complex is constructed by considering balls of radius $r$.
The sudden changes of persistent (sheaf) Betti number and the first nonzero eigenvalues near $r=0.65$\AA\ reflect the fact that most neighboring atoms are about 1.3\AA\ away from each other.
In Figure   \ref{mutation_nes_beta}, The number of atoms is reflected in the initial 0-th Betti numbers. The 0-th Betti number dramatically decreases around 0.65 \AA\ because covalent bond distances are about 1.5\AA. 
The 0-th Betti number decreases further from 1.2\AA\ to 1.7\AA\ due to other many non-covalent bonds.    

In Figure \ref{mutation_nes_lambda}, the results of the wild type and mutants almost coincide,  except that the first nonzero eigenvalues of persistent sheaf Laplacians of BA.1 and BA.2 near $r=0.65$\AA\ have very different values. 
The results of persistent Laplacians are quite different from those of persistent sheaf Laplacians at large filtration values. The significant changes around $r=0.65$\AA\ are due to the topological changes. 

We are also interested in understanding whether higher dimensional 
persistent Laplacians can offer an additional characterization of biomolecules. 
Figure \ref{miltidimen} presents the higher dimensional persistent Laplacian analysis of the wide type RBD near the N501 residue. Obviously, higher dimensional persistent Laplacian offers significant structural information about the distributions of circles and cavities of the macromolecule. Most dimension-1 circles occur in the range of 1.5-2.4\AA, whereas most 2-dimensional cavities locate around 1.8-2.8\AA. 
2-dimensional cavities are short-lived in the filtration, indicating the lack of multiple large cavities in the structure (at most one large cavity in the structure).  
This distribution can be used to understand interaction forces. For example, the length of hydrogen bonds ranges from 2-3.6\AA 
(corresponding to 1-1.8 \AA\ in the filtration radii).  This information is valuable for the design of machine learning representations, including the selection of the set of filtration intervals.   
{We also note that the peak of $\lambda_2^{r,0}$ is at the left of $\beta_2^{r,0}$. It's possible that when $r$ is in the range of 1.2\AA-1.5\AA, many 2-simplices are born but no 2-cycles are formed yet.}

\begin{figure}[htbp]
    \centering
        \includegraphics[width=0.9\linewidth]{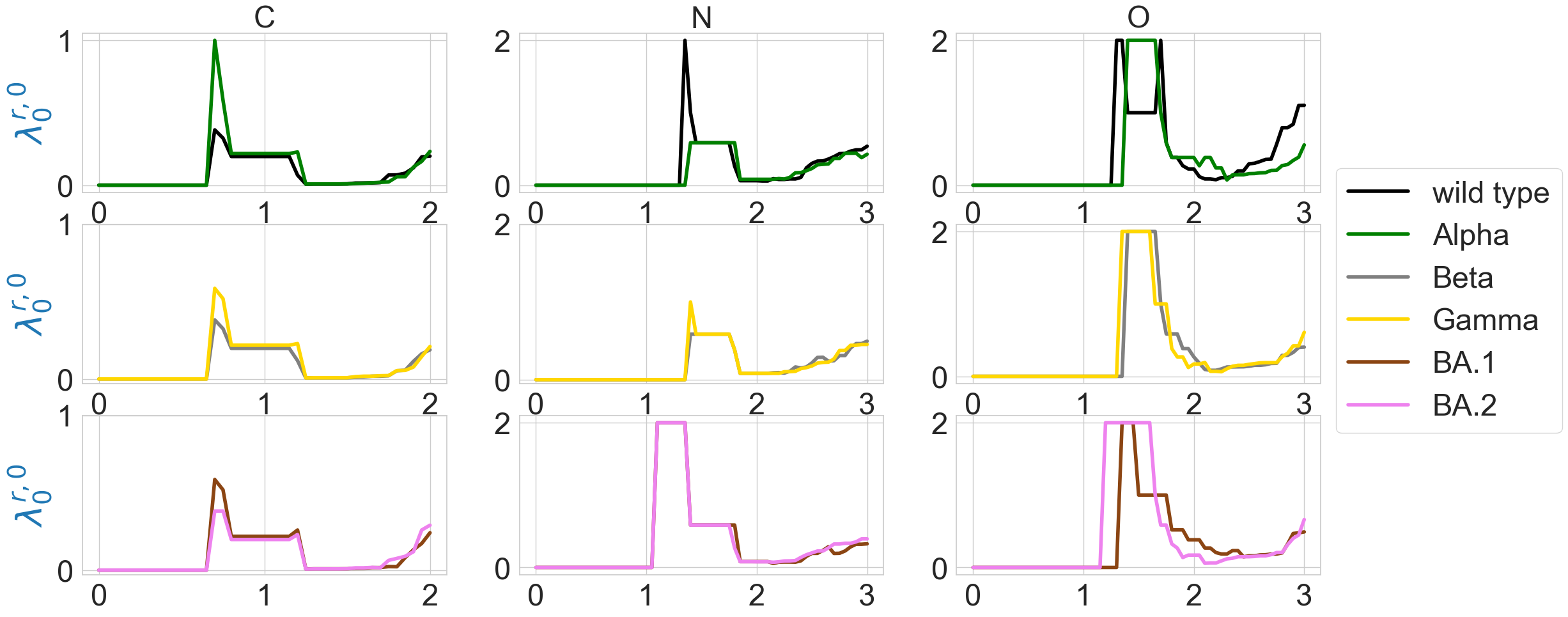}
    \caption{Illustration of the first nonzero eigenvalues of element-specific persistent Laplacian
    of the residue 501 mutation site at different filtration values, i.e., radii (unit: \AA). 
		The wild type (PDB ID: 6M0J) and Alpha (PDB ID: 8DLK) are given in the first row. 
		The Beta (PDB ID: 8DLN) and Gamma (PDB ID: 8DLQ) are given in the second row. 
		BA.1 (PDB ID: 7T9L) and BA.2 (PDB ID: 7XB0) are given the third row.
    }
    \label{mutation_es_lambda_pl}
\end{figure}

\begin{figure}[htbp]
    \centering
        \includegraphics[width=0.9\linewidth]{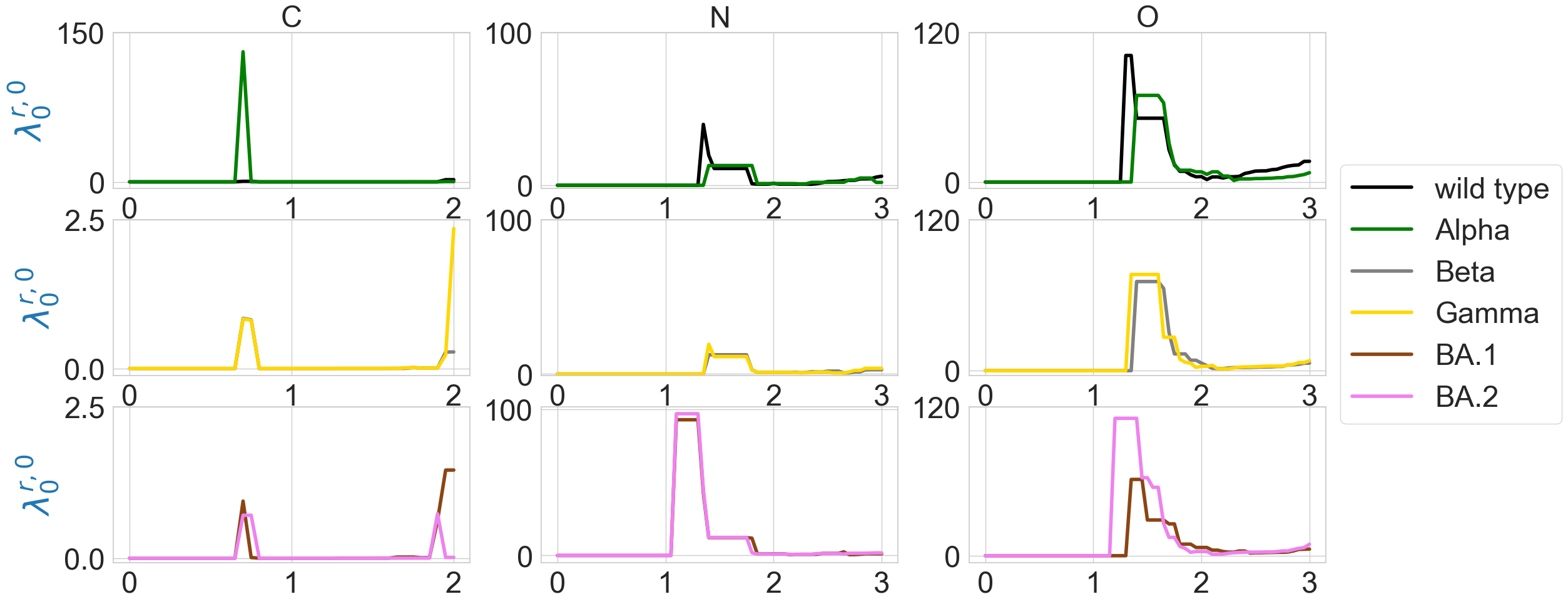}
    \caption{Illustration of the first nonzero eigenvalues  of element-specific persistent sheaf Laplacian
    of the residue 501 mutation site at different filtration values, i.e., radii (unit: \AA). 
		The wild type (PDB ID: 6M0J) and Alpha (PDB ID: 8DLK) are given in the first row. 
		The Beta (PDB ID: 8DLN) and Gamma (PDB ID: 8DLQ) are given in the second row. 
 BA.1 (PDB ID: 7T9L) and BA.2 (PDB ID: 7XB0) are given the third row. 
    }
    \label{mutation_es_lambda_psl}
\end{figure}

The element-specific results of the residue 501 mutation site of the wild type, and variants Alpha, Beta, Gamma, BA.1, and BA.2 are shown in 
Figures \ref{mutation_es_lambda_pl} and \ref{mutation_es_lambda_psl}, 
as well as in Figures \ref{mutation_es_beta_pl} and \ref{mutation_es_beta_psl} in the Appendix.
We observe that the difference between the first nonzero eigenvalues is much more obvious.
For instance, in Figure \ref{mutation_es_lambda_pl} there is a higher spike near 0.7\AA\ in the graph of Alpha carbon atoms, 
and two spikes near 1.3\AA\ and 1.7\AA\ disappear in the graph of the Alpha variant's oxygen atoms. 
In Figure \ref{mutation_es_lambda_pl}, all results of carbon atoms have similar shapes, implying a relatively stable RBD carbon atom structure.
In the results of nitrogen atoms, 
we notice that the results of Alpha, Beta, and Gamma variants resemble each other, 
and the same can be said of the results of BA.1 and BA.2 variants.
In the results of oxygen atoms, 
the results of Alpha, Beta, and Gamma still resemble each other, but the results of BA.1 and BA.2 are quite different. 
The results of the wild type are unique in the sense that it has one or two spikes near 1.3\AA\ or 1.7\AA.
These results indicate that element-specific persistent Laplacians and 
element-specific persistent sheaf Laplacians are better approaches in characterizing SARS-CoV-2 variants than element-non-specific approaches. 
{We know that nitrogen and oxygen atoms are sparser in a protein, 
so if we use element nonspecific approach, nitrogen atoms and oxygen atoms will first form edges with neighboring carbon atoms, 
and we are not able to infer distances between nitrogen atoms or oxygen atoms. This explains why element specific approach outperforms element nonspecific approach.}
 
\subsection{PTL analysis of RBD structural changes induced by its binding to ACE2}

We investigate how binding to ACE2 changes the spike protein RBD structure from the closed state to the open state for the wild type, Alpha, Beta, BA.1, and BA.2 variants. The PDB IDs of the spike protein of wild type, Alpha, Beta, BA.1 and BA.2 used in this section are 7DF3 \cite{xu2021conformational}, 7LWS \cite{gobeil2021effect}, 7LYM \cite{gobeil2021effect}, 7TF8 \cite{gobeil2022structural} and 7XIX \cite{cao2022ba}. 
The analysis of the Gamma variant is eliminated due to the lack of experimental structure.
We first align each of the three RBDs in the closed-state spike protein to the RBD in the RBD-ACE2 complex.
The maximal distances between corresponding atoms in the RBM of the three alignments of BA.1 are 8.76\AA, 13.49\AA, and 9.44\AA, which are larger than those of alignments of the wild type and other mutants.
For each alignment, we record the RBM residues that have at least one atom whose distance to the corresponding atom is larger than 5.28\AA, i.e., half of the mean maximal distances between corresponding atoms in RBM of the three alignments of BA.1.
In wild-type RBD, such residues are K444 and  K458. In Alpha there are no such residues;
In Beta, chains A and B have K458; chain C has T478 and P479. 
In BA.1, each chain has different such residues: chain A has K440, Y453, K458, K478, and F486;
chain B has K440, Y453, R457, K458, R466, Y473, Q474, K478, F486, F490, R493; and chain C has K440, Y453, Y473, K478, F486.
In BA.2 such residues are E465, K478, and G482.

\begin{figure}[htbp]
  \centering
  \includegraphics[width=0.8\linewidth]{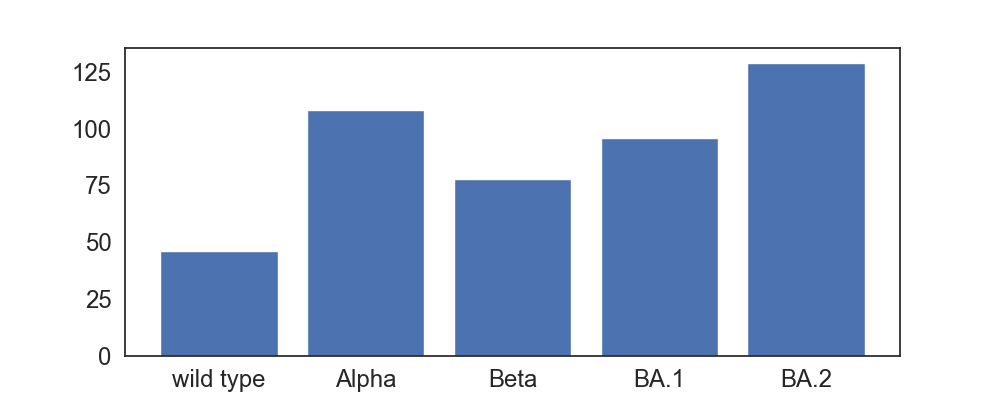}
  \caption{The total structural changes of the RBM between the closed state RBD and the open state RBD induced by ACE2 binding. 
  Here the total structural changes are defined to be the sum of squares of distances between C$_\alpha$ atoms in the RBM.
  }
  \label{total_structural_binding}
\end{figure}
We also calculate the total structural changes of the RBM between the closed state RBD and the open state RBD induced by its binding to the human ACE2. 
Here, the total structural changes are defined to be the sum of squares of distances between C$_\alpha$ atoms in the RBM.
Since spike protein is a trimer, we calculate the total structural changes for each chain and report the average (see Figure \ref{total_structural_binding}).
It turns out that the average total structural changes induced by binding to ACE2 do not increase too much with respect to the number of RBD mutations.

\begin{figure}[htbp]
  \centering
      \includegraphics[width=0.9\linewidth]{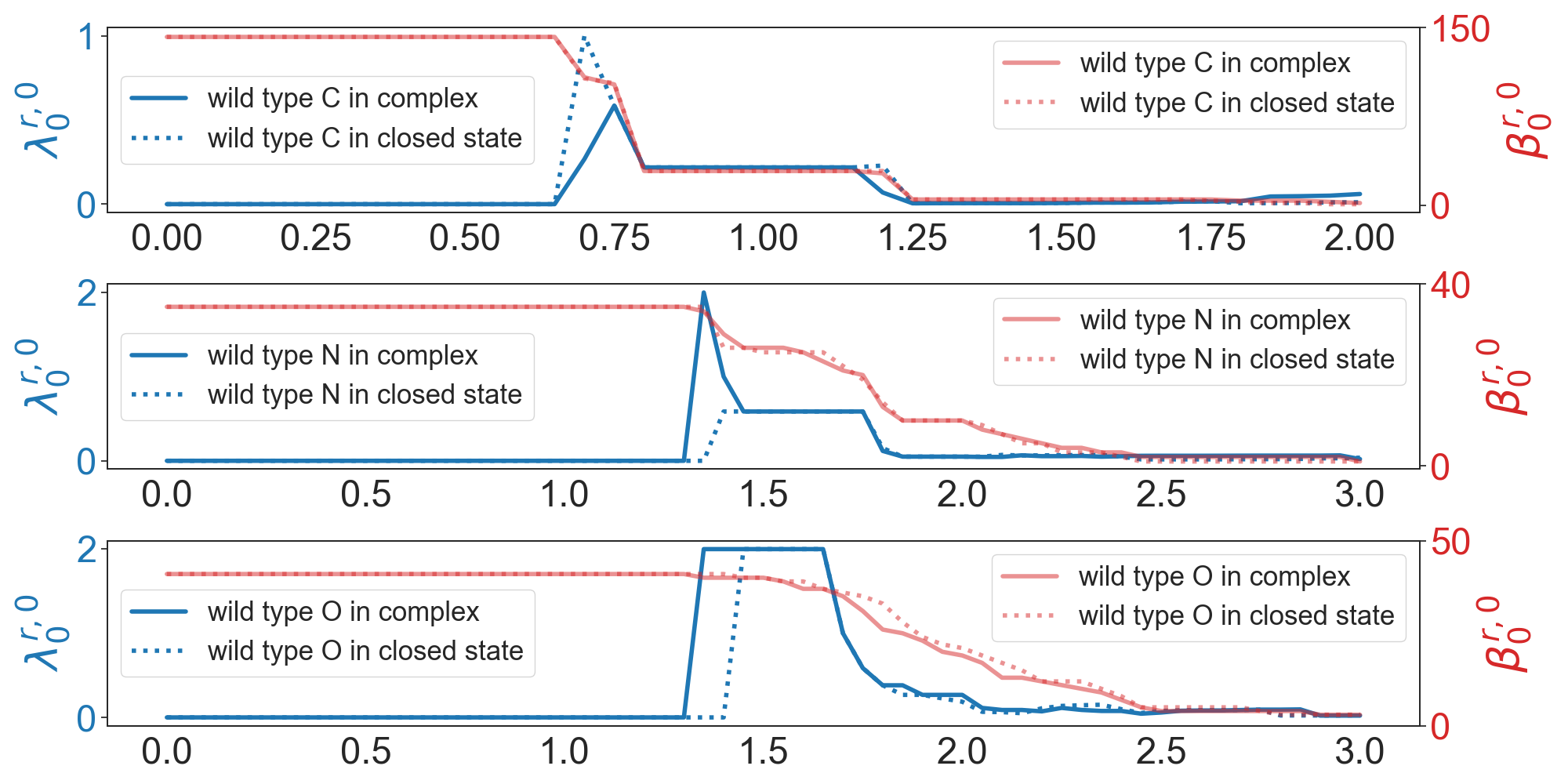}
  \caption{Illustration of persistent Betti numbers (red line) and the first nonzero eigenvalues (blue line) of persistent Laplacian
  of the RBD binding site of the wild type RBD-ACE2 complex (PDB ID: 6M0J) and closed state spike protein (PDB ID: 7DF3, Chain ID: A) at different filtration values, i.e., radii (unit: \AA). 
  The graphs from top to bottom represent the results of carbon atoms, nitrogen atoms, and oxygen atoms, respectively.}
  \label{bindedsurvsclosedsur_wild}
\end{figure}
Now, we calculate persistent Laplacians and persistent sheaf Laplacians for the RBD binding site in the closed state spike protein and the RBD-ACE2 complex.
For the wild type and mutants, we define the RBD binding site as the set of RBD residues whose C$_\alpha$s are within 10\AA\ from the C$_\alpha$s of ACE2 residues. 
We choose 10\AA\ as the cutoff distance, because if we used 11\AA\ then the RBD binding site would include non-RBM residues.
Spike protein as a trimer has three chains. 
In the results of alignments, the recorded residues of the wild type, Alpha, and BA.2 are the same for the three chains. Therefore,  for the wild type, Alpha and BA.2 we only use chain A, and for Beta and BA.1, we use all three chains. 
The study was carried out in an element-specific manner for carbon atoms, nitrogen atoms, and oxygen atoms. The results of the wild type are shown in Figure  \ref{bindedsurvsclosedsur_wild}.
We noted that persistent Betti numbers cannot distinguish two structures. However, the first nonzero eigenvalues of the persistent Laplacian capture the difference, demonstrating the advantage of persistent Laplacian over persistent homology in protein structure analysis.  

Additional analysis is presented in Figures \ref{bindedsur}, \ref{bindedsurvsclosedsur_alpha}, \ref{bindedsurvsclosedsur_beta}, \ref{bindedsurvsclosedsur_ba1}, \ref{bindedsurvsclosedsur_ba2}, \ref{bindedsurvsclosedsur_wild_psl}, \ref{bindedsurvsclosedsur_alpha_psl}, \ref{bindedsurvsclosedsur_beta_psl}, \ref{bindedsurvsclosedsur_ba1_psl}, and \ref{bindedsurvsclosedsur_ba2_psl} in the Appendix.
In Figure \ref{bindedsur}, the results of the wild type, Alpha, Beta, BA.1, and BA.2 RBD binding sites are quite similar except that the wild type RBD binding site has relatively lower first nonzero eigenvalues near $r=0.7$\AA.  
 It is seen that peak appears or disappears in the graph of the nitrogen atoms, whereas for BA.1 and BA.2, the results of the nitrogen atoms resemble each other, sometimes even coincide.

In general, the first nonzero eigenvalues of the persistent Laplacian are able to distinguish the structural difference before and after the complex formation in various variants. In contrast, the harmonic spectra, or equivalently, persistent homology, cannot always capture the structural changes.   

The results of persistent Laplacians and persistent sheaf Laplacians are similar in this work. However, this similarity is due to the specific implementation of persistent sheaf Laplacians. In general,  persistent sheaf Laplacians enable the embedding of non-geometric chemical and physical information of biomeolecules in topological and spectral representations.

\subsection{Impacts of computationally generated mutant structures on PTL-based topological deep learning predictions}

\begin{figure}[htbp]
    \centering
    \includegraphics[width=0.3\linewidth]{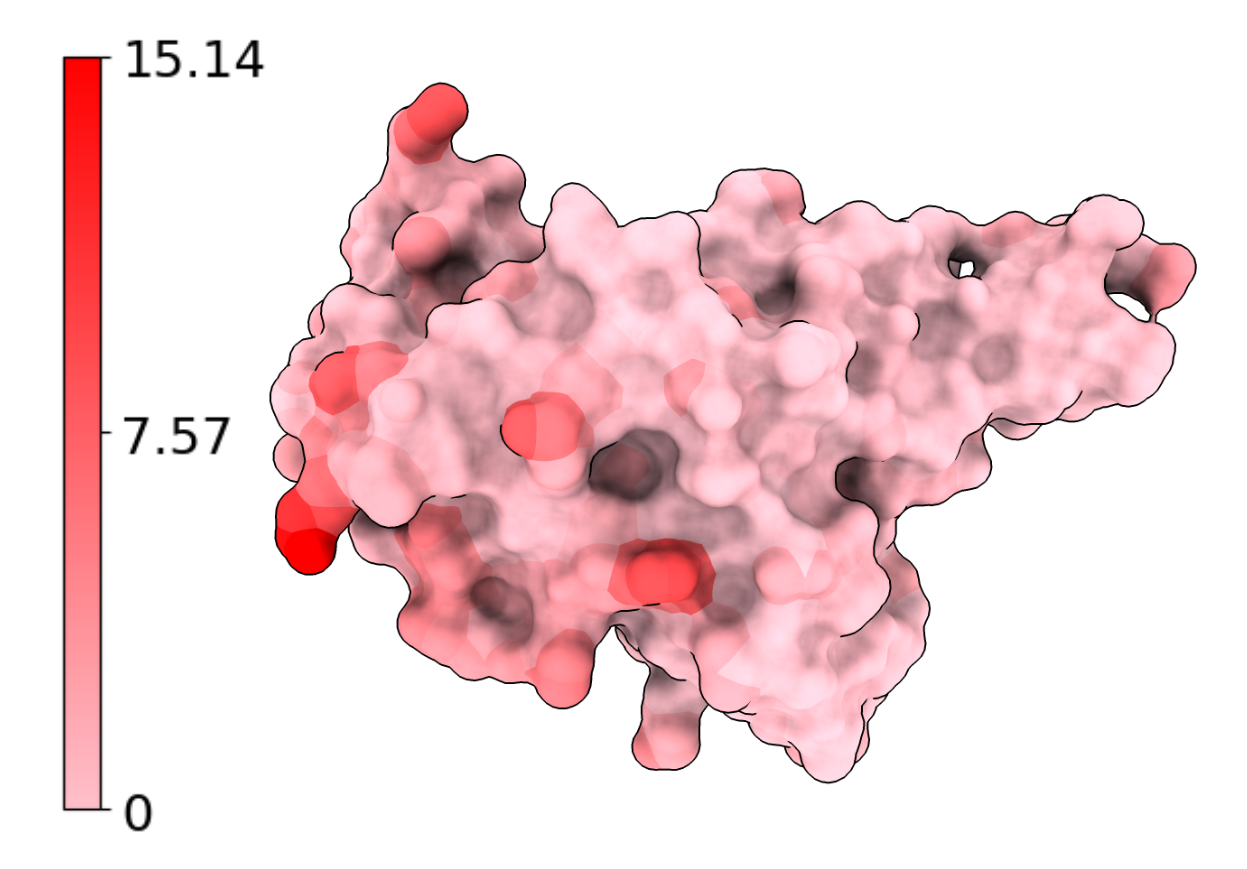}
    \caption{Atoms of BA.2 RBD (PDB ID: 7XB0) are colored by their distances to corresponding atoms in the computationally generated structure.
    Blue, white, and red corresponds to 0\AA, 7.57\AA, and 15.14\AA\, respectively. 
    We record the residues that have at least one atom whose distance to the corresponding atom in wild type RBD is more than 7.57\AA.
    Such residues are 370, 375, 378, 386, 387, and 519. 
    }
    \label{alignment_comvsexp}
\end{figure}

 The understanding of PPIs is a vital task in computational biology. With the availability of large amounts of good quality data,  machine learning approaches have demonstrated their unique capability \cite{liu2022hom}. This is specifically true for the prediction of SARS-CoV-2 infectivity and antibody resistance using topological deep learning \cite{chen2022omicron}. 
 In this work, we explore the impact of computationally generated structures on the predictive accuracy of our topological deep learning framework.  
We use a BA.2 RBD deep mutational scanning dataset which involves the systematical mutations of each residue on the BA.2 RBD to 19 others and records corresponding binding affinity changes \cite{starr2022deep}. 

 The deep mutational scanning covers the RBD residues from 333 to 527. 
In order to apply machine learning models, such as TopLapGBT and TopLapNet \cite{chen2022toplapnet}, to this dataset, BA.2 RBD mutants need to be computationally generated based on a BA.2 RBD structure and the choice of the BA.2 RBD structure can affect the performance of machine learning models.
We can employ either an experimentally determined BA.2 RBD-ACE2 complex structure or a BA.2 RBD-ACE2 complex structure computationally generated based on an experimentally determined BA.1 RBD-ACE2 complex structure. 
These two complexes are systematically mutated to all possible mutants in the deep mutational scanning dataset \cite{starr2022deep}. Two deep learning models, namely TopLapGBT and TopLapNet, are used to predict the binding affinity changes induced by all  BA.2 RBD mutations.  The results from these complexes are compared to examine the performance of computationally generated mutants. Here, we computationally generate mutant structures for mutations L371F, T376A, D405N, R408S, S446G, and S496G.

When the given BA.2 RBD structure is experimentally determined (PDB ID: 7XB0),
and the resulting models are referred to as ExpTopLapGBT (experimental TopLapGBT) and ExpTopLapNet. 
When the BA.2 RBD structure is computationally generated from BA.1 RBD (PDB ID: 7T9L) by Jackal \cite{xiang2001extending}, 
the resulting model is referred to as ComTopLapGBT (computational TopLapGBT) or ComTopLapNet.
The distances of corresponding atoms between the experimentally determined RBD (PDB ID: 7XB0) and the RBD generated computationally from BA.1 RBD (PDB ID: 7T9L) is shown in Figure \ref{alignment_comvsexp}. 

\begin{center}
    \begin{tabular}{ |c|c|c|c| } 
     \hline
     Method & $R_p(Exp, True)$ & $R_p(Com, True)$ & $R_p(Exp, Com)$ \\ 
     \hline
    TopLapGBT  & 0.901 & 0.898 & 0.990 \\ 
     \hline
    TopLapNet  & 0.879 & 0.849 & 0.925 \\ 
     \hline
    \end{tabular}
    \captionof{table}{  $R_p(Exp, True)$ is the correlation coefficient between predictions of ExpTopLapGBT (or ExpTopLapNet) and true affinity changes. Here, $R_p(Com, True)$ is the correlation coefficient between predictions of ComTopLapGBT (or ComTopLapNet) and true affinity changes.
    $R_p(Exp, Com)$ is Pearson the correlation coefficient between the predictions of ExpTopLapGBT and ComTopLapGBT (or between ExpTopLapNet and ComTopLapNet). A random state affects the 10-fold splitting and the training of GBT and neural networks.
    }
    \label{toplapgbt}
\end{center}

To evaluate the validity of computationally generated BA.2 complex structure, we compare the results of two topological deep learning methods, ExpTopLapGBT and ComTopLapGBT, on the predictions of the RBD deep mutational scanning dataset.   
We split the dataset into 10 folds, and for each fold, we use the other 9 folds as the training set to build a machine learning model, which is used to predict ACE2-binding affinity changes for the fold.
Therefore,  for a given 10-fold splitting we get the  ExpTopLapGBT and ComTopLapGBT predictions of RBD-ACE2 binding affinity changes for the deep mutational scanning dataset. 
We denote $R_p(Exp, True)$ the Pearson correlation coefficient between ExpTopLapGBT predicted binding affinity changes and experimental binding affinity changes. Similarly, $R_p(Com, True)$ (or $R_p(Exp, Com)$) is the Pearson correlation coefficient between ComTopLapGBT predicted binding affinity changes and experimental binding affinity changes (or ExpTopLapGBT predicted binding affinity changes).

The results of TopLapGBT and TopLapNet are shown in Table  \ref{toplapgbt}. 
Generally, the performance of models using experimentally determined structures is better than that of models using the computationally generated structure. 
This is not surprising since the computationally generated structure is an approximation of the experimental structure. 
The performance of ExpTopLapGBT and ComTopLapGBT are extremely close,
whereas the performance of ComTopLapNet differs very much from that of ExpTopLapNet. 
We also see that  ExpTopLapGBT outperforms  ExpTopLapNet.

\section{Theories and methods} \label{}

\subsection{Persistent topological Laplacians} 

Persistent topological Laplacians (PTLs) are a family of topological data analysis methods that are topological, multiscale, and spectral. Loosely speaking, their kernel space dimensions coincide with the topological invariants or Betti numbers in each topological dimension and their non-harmonic spectra describe homotopic shape evolution during filtration or multiscale analysis. Various discrete PTLs,  e.g., persistent Laplacian \cite{wang2020persistent},  persistent sheaf Laplacian \cite{wei2021persistent},   persistent path Laplacian \cite{wang2023persistent}, and persistent directed hypergraph Laplacian \cite{chen2023persistent} have been proposed for point cloud data. For  volumetric data, evolutionary de Rham-Hodge method has been developed \cite{chen2021evolutionary}, which  is defined on a family of evolving manifolds. Evolutionary de Rham-Hodge method is based on differential geometry, algebraic topology, multiscale analysis, and partial differential equations. In this work, we focus on persistent Laplacian and  persistent sheaf Laplacian.

Suppose $K$ and $L$ are two simplicial complexes and $K$ is a subset of $L$. We denote by $C^K$ and $C^L$ the simplicial chain complexes of $K$ and $L$ with real coefficients.
As a chain group $C_q$ in a simplicial chain complex is formally generated by simplices, 
it is naturally a finite-dimensional inner product space, and the adjoint of boundary map $\partial_q$ is well defined.
Let $C_{q+1}^{L,K}$ be the subspace $\{c \in C_{q+1}^L \mid \partial_{q+1}^L(c) \in C_q^K \}$ of $C_{q+1}^L$ 
and $\partial_{q+1}^{L,K}$ the restriction of $\partial^L_{q+1}$ to $C_{q+1}^{L,K}$.
The $q$-th persistent Laplacian $\Delta_{q}^{K,L}$ is defined by 
\begin{align}
    \partial_{q+1}^{L,K} (\partial_{q+1}^{L,K})^{\ast} + (\partial^K_{q})^{\ast} \partial^K_{q}.
\end{align}

Before we define the persistent sheaf Laplacian, we need to explain what a cellular sheaf is first.
A cellular sheaf $\mathscr{S}$ is a simplicial complex $X$ (viewed as a cell complex) with 
an assignment to each cell $\sigma$ of $X$ a finite-dimensional vector space $\mathscr{S}(\sigma)$ (referred to as the stalk of $\mathscr{S}$ over $\sigma$)
and to each face relation $\sigma \leqslant \tau$ (i.e.,\ $\sigma \subset \overline{\tau}$)
a linear morphism of vector spaces denoted by
$\mathscr{S}_{\sigma \leqslant \tau}$ (referred to as the restriction map of the face relation $\sigma \leqslant \tau$), satisfying the rule 
\begin{align*}
    \rho \leqslant \sigma \leqslant \tau \Rightarrow \mathscr{S}_{\rho \leqslant \tau} = \mathscr{S}_{\sigma \leqslant \tau} \mathscr{S}_{\rho \leqslant \sigma}
\end{align*}
and $\mathscr{S}_{\sigma \leqslant \sigma}$ is the identity map of $\mathscr{S}(\sigma)$. 
Like a simplicial complex, a cellular sheaf gives rise to a sheaf cochain complex.
The $q$-th sheaf cochain group $C^q_\mathscr{S}$ is the direct sum of stalks over $q$-dimensional cells.
To define coboundary maps, we can globally orient the simplicial complex $X$ and obtain a signed incidence relation, i.e.\ an assignment to each $\sigma \leqslant \tau$ an integer $[\sigma: \tau]$. 
Then the coboundary map $d^{q}: C^q_\mathscr{S} \to C^{q+1}_\mathscr{S}$ is defined by 
\begin{align*}
    d^q \vert_{\mathscr{S}(\sigma)} = \sum_{\sigma \leqslant \tau} [\sigma: \tau] \mathscr{S}_{\sigma \leqslant \tau}.
\end{align*} 
Now suppose we have two cellular sheaves $\mathscr{S}$ on $K$ and $\mathscr{T}$ on $L$ such that 
$K \subseteq L$ and stalks and restriction maps of $K$ are identical to those of $L$.
\[
\begin{tikzcd}
    C^{q-1}_{\mathscr{S}} \arrow[rr, "d^{q-1}_{\mathscr{S}}", shift left] 
     &
      & C^{q}_{\mathscr{S}} \arrow[rr, "d^q_{\mathscr{S}}"] \arrow[ll, "(d^{q-1}_{\mathscr{S}})^{\ast}", shift left] \arrow[rd, "d^{q}_{\mathscr{S},\mathscr{T}}", shift left]
       & 
        & C^{q+1}_{\mathscr{S}}  \\
     & 
      &
       & C^{q+1}_{\mathscr{S}, \mathscr{T}} \arrow[lu, "(d^{q}_{\mathscr{S},\mathscr{T}})^{\ast}", shift left, pos=0.1] \arrow[rd, hook, dashed]
        & \\
    C^{q-1}_{\mathscr{T}} \arrow[uu, "\pi"] \arrow[rr, "d^{q-1}_{\mathscr{T}}"] 
     &
      & C^q_{\mathscr{T}} \arrow[rr, "d^q_{\mathscr{T}}"] \arrow[uu, "\pi"] 
       & 
        & C^{q+1}_{\mathscr{T}} \arrow[uu, "\pi"]
\end{tikzcd}
\]
Let $C_{\mathscr{S},\mathscr{T}}^{q+1} = \{c \in C^{q+1}_\mathscr{T} \mid (d^{q}_{\mathscr{T}})^{\ast}(c) \in C^{q}_\mathscr{S} \}$.
We denote the adjoint map of $(d^{q}_\mathscr{T})^{\ast}\vert_{C_{\mathscr{S},\mathscr{T}}^{q+1}}$ as $d^{q}_{\mathscr{S},\mathscr{T}}$
and define the $q$-th persistent sheaf Laplacian $\Delta_q^{\mathscr{S},\mathscr{T}}$ as
\begin{align*}
    \Delta_q^{\mathscr{S},\mathscr{T}} = (d^q_{\mathscr{S},\mathscr{T}})^{\ast} d^q_{\mathscr{S},\mathscr{T}} + d^{q-1}_{\mathscr{S}} (d^{q-1}_{\mathscr{S}})^{\ast}.
\end{align*}
 Of course, to define adjoint maps cochain groups need to be inner product spaces.
In this work, cellular sheaves are constructed in the same way as in Section 2.4 of \cite{wei2021persistent}.
  The non-geometrical information we consider is the set of atomic partial charges. 
We employ  partial charges from the PDB2PQR package \cite{jurrus2018improvements}. 
We build a Rips filtration of graphs. 
For each  simplicial complex $X$, we denote each vertex by $v_i$, the edge connecting $v_i$ and $v_j$ by $e_{ij}$ and the partial charges $q_i$.
Then the cellular sheaf is such that each stalk is $\mathbb{R}$, and for face relation $v_i \leqslant e_{ij}$ the morphism is 
the multiplication by $q_j/r_{ij}$, where $r_{ij}$ is the length of $e_{ij}$. 
Spectra of persistent sheaf Laplacians are not used in TopLapGBT and TopLapNet.

\subsection{TopLapGBT and TopLapNet} \label{toplap}

TopLapGBT and TopLapNet \cite{chen2022toplapnet} have been employed to study mutational effects on protein-protein interaction. Gradient boosted trees are employed in TopLapGBT, whereas,   TopLapNet is based on artificial neural networks. Both models are constructed by using persistent Laplacians. These methods require the 3D structures of both wide-type PPI complexes and  mutant complexes.  
For instance, the AB-Bind S645 dataset\cite{sirin2016ab} includes 645 mutants with experimentally determined BFE changes across 29 antibody-antigen complexes.
Mutant structures can be computationally generated based on experimentally determined structures of the wild-type antibody-antigen complexes and mutation information (chain id, residue id, mutant residue, etc.).
Representation of mutant structures, including the persistent homology and persistent Laplacian representations, and other auxiliary representations, 
can be used as feature vectors to train machine learning models (such as gradient boosting trees and deep neural networks)
that can predict mutation-induced BFE changes.

When we apply persistent homology and persistent Laplacian to the study of protein-protein interactions,
we always extract the atoms within a certain cutoff distance $r$ of the binding site 
and construct a distance matrix such that if two atoms are in the same protein then the distance between them is an extremely large constant number. 
If we want to further characterize the interaction between atoms of certain elements $E_1$ and $E_2$, 
we can consider the point cloud formed by the atoms of an element $E_1$ of protein $A$ within $r$ of the binding site,
and the atoms of element $E_2$ of protein $B$ within $r$ of the binding site.
After the calculation of persistent homology and persistent Laplacian, 
the next step is to transform the barcodes of persistent homology or spectra of persistent Laplacians into vector representations of fixed lengths. 
For barcodes, there are at least two ways: 
either we divide the interval $[0, r]$ into bins of even length
and count the occurrence of bars, birth values, and death values in each bin, 
or we simply compute statistics such as sum, maximum, minimum, mean, and standard deviation for bar lengths, birth values, and death values.
The former method is often applied to 0-dimensional barcodes and the latter to 1-dimensional and 2-dimensional barcodes.
For the spectrum of a persistent Laplacian, we separate zero eigenvalues (harmonic spectra) and nonzero eigenvalues (non-harmonic spectra).
We use the number of zero eigenvalues, the sum, the minimum, the maximum, the mean, the standard deviation, the variance, and the sum of squares of nonzero eigenvalues. 

In this study, we use scikit-learn to build a gradient boosting tree whose parameters are n\_estimators=20000, learning\_rate = 0.005, max\_features = `sqrt', max\_depth = 9, min\_samples\_split = 3, subsample = 0.4, and n\_iter\_no\_change=500. Additionally, we use PyTorch to build a neural network with 7 hidden layers and each layer has 8000 neurons.

\section{Concluding remarks}

Persistent topological Laplacians (PTLs) are a class of newly proposed multiscale topological spectral approaches in data science. These methods can be used either in a discrete  setting for point cloud data  \cite{wang2020persistent, wei2021persistent, wang2023persistent} or in a continuous setting for volumetric data \cite{chen2021evolutionary}. 
Their mathematical underpinnings for discrete formulations are algebraic topology, sheaf theory, and combinatorial graphs, while those for the continuous formulations is algebraic topology, differential geometry, and partial differential equation. 
Among mutants Alpha, Beta, Gamma, Omicron BA.1, and Omicron BA.2, BA.2 has the largest total structural changes from the wild type, 
which agrees with the significant antibody escape of the Omicron BA.2 variant.
As to the total structural changes of a closed state RBD induced by its binding to ACE2,
total structural changes of Alpha, Beta, Omicron BA.1, and Omicron BA.2 do not differ too much.
It is noted that most large structural changes of C$_\alpha$ occur at flexible random coil regions at the epitope.     

We also demonstrate how to use PTLs to characterize structural changes induced by SARS-CoV-2 variant spike protein receptor-binding domain (RBD) mutations and by its binding to human angiotensin-converting enzyme 2 (ACE2). Two PTLs, namely persistent   Laplacian and persistent sheaf Laplacian, are utilized in our work. 
We also analyze two implementations, i.e., element-nonspecific and element-specific Laplacian models of persistent Laplacian and persistent sheaf Laplacian. 
We show that persistent Laplacian and persistent sheaf Laplacian provide similar results. These methods capture homotopic shape evolution information, which persistent homology cannot offer. We expect other persistent topological Laplacians, such as persistent path Laplacian \cite{wang2023persistent}, can uncover similar information.   Additionally, element-specific approaches reveal more information than element-nonspecific ones as shown in literature \cite{cang2017topologynet}. 
  
More specifically,  the results of persistent Laplacian and persistent sheaf Laplacian indicate that at the residue 501 mutation site,
the structure of RBD carbon atoms is less affected by mutations than that of RBD nitrogen atoms and RBD oxygen atoms, partially due to the fact that the bond nature of carbon atoms is mostly covalent, whereas the distances among oxygen atoms are mostly non-covalent. The topological Betti numbers of oxygen atoms are associated with possible hydrogen bonds.     
Additionally, structural similarity of mutation sites is observed in the spectra of persistent Laplacian and persistent sheaf Laplacian.

As for the RBD structural changes induced by its binding to ACE2, for wild type, Alpha and Beta, a significant difference can be observed in the result of nitrogen atoms, 
whereas for BA.1 and BA.2, a significant difference can be observed in the result of oxygen atoms. We show that the non-harmonic spectra (the first non-zero eigenvalues) of persistent Laplacian are more sensitive to structural changes than the harmonic spectra.  Therefore, persistent Laplacian has an advantage over persistent homology in protein analysis. 

Finally we test how a computationally generated structure impacts 
the prediction of PTL-based machine learning models, i.e., TopLapGBT and TopLapNet \cite{chen2022toplapnet}. 
The results indicate that a computationally generated structure harms the performance of TopLapGBT and TopLapNet. 
However, TopLapGBT is much less affected by a computationally generated structure than TopLapNet, 
implying a resistance to the structural approximation from computations.
TopLapNet is more affected by a computationally generated structure probably because neural networks are more prone to overfitting than gradient-boosted trees. 

This work reveals that PTLs are a class of powerful new methods for topological data analysis (TDA) or more precisely, spectral data analysis (SDA). These methods can certainly be applied to the data analysis in other fields and disciplines, including image science, physical science, medical science,  social science, engineering, financial industrial, musical science \cite{wei2023topological}, etc.

	\section*{Data  availability}
		
		The related datasets studied in this work are available at: 
		https://github.com/WeilabMSU/PTLvirus.

\section*{Acknowledgment}
This work was supported in part by NIH grants  R01GM126189 and  R01AI164266, NSF grants DMS-2052983,  DMS-1761320, and IIS-1900473,  NASA grant 80NSSC21M0023,  MSU Foundation,  Bristol-Myers Squibb 65109, and Pfizer.

\vspace{1cm} 

\begin{thebibliography}{10}

\bibitem{ameneyro2022quantum}
B.~Ameneyro, G.~Siopsis, and V.~Maroulas.
\newblock Quantum Persistent Homology.
\newblock {arXiv preprint arXiv:2202.12965}, 2022.

\bibitem{cang2018representability}
Z.~Cang, L.~Mu, and G.-W. Wei.
\newblock Representability of algebraic topology for biomolecules in machine
  learning based scoring and virtual screening.
\newblock {\em PLoS computational biology}, 14(1):e1005929, 2018.

\bibitem{cang2017topologynet}
Z.~Cang and G.-W. Wei.
\newblock TopologyNet: Topology based deep convolutional and multi-task neural
  networks for biomolecular property predictions.
\newblock {\em PLoS Computational Biology}, 13(7):e1005690, 2017.

\bibitem{cao2022ba}
Y.~Cao, A.~Yisimayi, F.~Jian, W.~Song, T.~Xiao, L.~Wang, S.~Du, J.~Wang, Q.~Li,
  X.~Chen, et~al.
\newblock BA.2.12.1, BA.4 and BA.5 escape antibodies elicited by omicron
  infection.
\newblock {\em Nature}, 608(7923):593--602, 2022.

\bibitem{carlsson2009topology}
G.~Carlsson.
\newblock Topology and data.
\newblock {\em Bulletin of the American Mathematical Society}, 46(2):255--308,
  2009.

\bibitem{chen2023persistent}
D.~Chen, J.~Liu, J.~Wu, and G.-W. Wei.
\newblock Persistent topological hypergraph and hyperdigraph Laplacians.
\newblock {\em in preparation}, 2023.

\bibitem{chen2021prediction}
J.~Chen, K.~Gao, R.~Wang, and G.-W. Wei.
\newblock Prediction and mitigation of mutation threats to {COVID-19} vaccines
  and antibody therapies.
\newblock {\em Chemical Science}, 12(20):6929--6948, 2021.



\bibitem{chen2021revealing}
J.~Chen, K.~Gao, R.~Wang, and G.-W. Wei.
\newblock Revealing the threat of emerging {SARS-CoV-2} mutations to antibody
  therapies.
\newblock {\em Journal of Molecular Biology}, 433(7744), 2021.

\bibitem{chen2022toplapnet}
J.~Chen, Y.~Qiu, R.~Wang, and G.-W. Wei.
\newblock Persistent Laplacian projected omicron BA.4 and BA.5 to become new
  dominating variants.
\newblock {\em Computers in Biology and Medicine}, 151, 106262, 2022.

\bibitem{chen2022omicron}
J.~Chen, R.~Wang, N.~B. Gilby, and G.-W. Wei.
\newblock Omicron variant (B.1.1.529): infectivity, vaccine breakthrough, and
  antibody resistance.
\newblock {\em Journal of Chemical Information and Modeling}, 62(2):412--422,
  2022.

\bibitem{chen2020mutations}
J.~Chen, R.~Wang, M.~Wang, and G.-W. Wei.
\newblock Mutations strengthened SARS-CoV-2 infectivity.
\newblock {\em Journal of Molecular Biology}, 432(19):5212--5226, 2020.

\bibitem{chen2022omicron2}
J.~Chen  and G.-W. Wei.
\newblock Omicron BA. 2 (B. 1.1. 529.2): high potential for becoming the next dominant variant.
\newblock {\em The Journal of Physical Chemistry Letters}, 13(17):3840--3849,
  2022.

\bibitem{chen2021evolutionary}
J.~Chen, R.~Zhao, Y.~Tong, and G.-W. Wei.
\newblock Evolutionary de Rham-Hodge method.
\newblock {\em Discrete and Continuous Dynamical Systems. Series B},
  26(7):3785, 2021.

\bibitem{edelsbrunner2000topological}
H.~Edelsbrunner, D.~Letscher, and A.~Zomorodian.
\newblock Topological persistence and simplification.
\newblock In {\em Proceedings 41st Annual Symposium on Foundations of Computer
  Science}, pages 454--463. IEEE, 2000.

\bibitem{frosini1992measuring}
P.~Frosini.
\newblock Measuring shapes by size functions.
\newblock In {\em Intelligent Robots and Computer Vision X: Algorithms and
  Techniques}, volume 1607, pages 122--133. International Society for Optics
  and Photonics, 1992.

\bibitem{gobeil2022structural}
S.~M.-C. Gobeil, R.~Henderson, V.~Stalls, K.~Janowska, X.~Huang, A.~May,
  M.~Speakman, E.~Beaudoin, K.~Manne, D.~Li, et~al.
\newblock Structural diversity of the SARS-CoV-2 omicron spike.
\newblock {\em Molecular Cell}, 2022.

\bibitem{gobeil2021effect}
S.~M.-C. Gobeil, K.~Janowska, S.~McDowell, K.~Mansouri, R.~Parks, V.~Stalls,
  M.~F. Kopp, K.~Manne, D.~Li, K.~Wiehe, et~al.
\newblock Effect of natural mutations of SARS-CoV-2 on spike structure,
  conformation, and antigenicity.
\newblock {\em Science}, 373(6555):eabi6226, 2021.

\bibitem{grbic2022aspects}
J.~Grbic, J.~Wu, K.~Xia, and G.~Wei.
\newblock Aspects of topological approaches for data science.
\newblock {\em Foundations of Data Science}, 2022.

\bibitem{hansen2019toward}
J.~Hansen and R.~Ghrist.
\newblock Toward a spectral theory of cellular sheaves.
\newblock {\em Journal of Applied and Computational Topology}, 3(4):315--358,
  2019.

\bibitem{hoffmann2020sars}
M.~Hoffmann, H.~Kleine-Weber, S.~Schroeder, N.~Kr{\"u}ger, T.~Herrler,
  S.~Erichsen, T.~S. Schiergens, G.~Herrler, N.-H. Wu, A.~Nitsche, et~al.
\newblock {SARS-CoV-2} cell entry depends on {ACE2} and {TMPRSS2} and is
  blocked by a clinically proven protease inhibitor.
\newblock {\em Cell}, 181(2):271--280, 2020.


\bibitem{jurrus2018improvements}
Jurrus, Elizabeth and Engel, Dave and Star, Keith and Monson, Kyle and Brandi, Juan and Felberg, Lisa E and Brookes, David H and Wilson, Leighton and Chen, Jiahui and Liles, Karina and others
\newblock Improvements to the APBS biomolecular solvation software suite.
\newblock {\em Protein Science}, 27(1):112--128, 2018.



\bibitem{kovacev2016using}
V.~Kovacev-Nikolic, P.~Bubenik, D.~Nikoli{\'c}, and G.~Heo.
\newblock Using persistent homology and dynamical distances to analyze protein
  binding.
\newblock {\em Statistical Applications in Genetics and Molecular Biology},
  15(1):19--38, 2016.

\bibitem{lan2020structure}
J.~Lan, J.~Ge, J.~Yu, S.~Shan, H.~Zhou, S.~Fan, Q.~Zhang, X.~Shi, Q.~Wang,
  L.~Zhang, et~al.
\newblock Structure of the SARS-CoV-2 spike receptor-binding domain bound to
  the ACE2 receptor.
\newblock {\em Nature}, 581(7807):215--220, 2020.

\bibitem{li2022structural}
L.~Li, H.~Liao, Y.~Meng, W.~Li, P.~Han, K.~Liu, Q.~Wang, D.~Li, Y.~Zhang,
  L.~Wang, et~al.
\newblock Structural basis of human ACE2 higher binding affinity to currently
  circulating Omicron SARS-CoV-2 sub-variants BA. 2 and BA. 1.1.
\newblock {\em Cell}, 185(16):2952--2960, 2022.

\bibitem{liu2023neighborhood}
J.~Liu, D. ~ Chen, F. ~Pan, and J. ~Wu.
\newblock Neighborhood Path Complex for the Quantitative Analysis of the Structure and Stability of Carboranes.
\newblock {\em Journal of Computational Biophysics and Chemistry}, https://doi.org/10.1142/S2737416523500229,
2023.

\bibitem{liu2022biomolecular}
J.~Liu, K.-L. Xia, J.~Wu, S.~S.-T. Yau, and G.-W. Wei.
\newblock Biomolecular topology: Modelling and analysis.
\newblock {\em Acta Mathematica Sinica, English Series}, 38(10):1901--1938,
  2022.

\bibitem{liu2022hom}
X.~Liu, H.~Feng, J.~Wu, and K.~Xia.
\newblock Hom-complex-based machine learning (HCMl) for the prediction of
  protein--protein binding affinity changes upon mutation.
\newblock {\em Journal of Chemical Information and Modeling},
  62(17):3961--3969, 2022.

\bibitem{mannar2022sars}
D.~Mannar, J.~W. Saville, Z.~Sun, X.~Zhu, M.~M. Marti, S.~S. Srivastava, A.~M.
  Berezuk, S.~Zhou, K.~S. Tuttle, M.~D. Sobolewski, et~al.
\newblock SARS-CoV-2 variants of concern: spike protein mutational analysis and
  epitope for broad neutralization.
\newblock {\em Nature Communications}, 13(1):1--12, 2022.

\bibitem{mannar2022omicron}
D.~Mannar, J.~W. Saville, X.~Zhu, S.~S. Srivastava, A.~M. Berezuk, K.~S.
  Tuttle, A.~C. Marquez, I.~Sekirov, and S.~Subramaniam.
\newblock SARS-CoV-2 Omicron variant: Antibody evasion and cryo-EM structure of
  spike protein--ACE2 complex.
\newblock {\em Science}, 375(6582):760--764, 2022.

\bibitem{memoli2022persistent}
F.~M{\'e}moli, Z.~Wan, and Y.~Wang.
\newblock Persistent Laplacians: Properties, algorithms and implications.
\newblock {\em SIAM Journal on Mathematics of Data Science}, 4(2):858--884,
  2022.

\bibitem{meng2021persistent}
Z.~Meng and K.~Xia.
\newblock Persistent spectral--based machine learning (perspect ml) for
  protein-ligand binding affinity prediction.
\newblock {\em Science Advances}, 7(19):eabc5329, 2021.

\bibitem{nguyen2019mathematical}
D.~D. Nguyen, Z.~Cang, K.~Wu, M.~Wang, Y.~Cao, and G.-W. Wei.
\newblock Mathematical deep learning for pose and binding affinity prediction
  and ranking in D3R grand challenges.
\newblock {\em Journal of Computer-Aided Molecular Design}, 33(1):71--82, 2019.

\bibitem{pettersen2021ucsf}
E.~F. Pettersen, T.~D. Goddard, C.~C. Huang, E.~C. Meng, G.~S. Couch, T.~I.
  Croll, J.~H. Morris, and T.~E. Ferrin.
\newblock UCSF  Chimerax: Structure visualization for researchers, educators,
  and developers.
\newblock {\em Protein Science}, 30(1):70--82, 2021.

\bibitem{qiu2022persistent}
Y.~Qiu and G.-W. Wei.
\newblock Persistent spectral theory-guided protein engineering.
\newblock {\em Nature Computational Science}, 2023.

\bibitem{sirin2016ab}
S.~Sirin, J.~R. Apgar, E.~M. Bennett, and A.~E. Keating.
\newblock {AB-Bind}: antibody binding mutational database for computational
  affinity predictions.
\newblock {\em Protein Science}, 25(2):393--409, 2016.

\bibitem{starr2022deep}
T.~N. Starr, A.~J. Greaney, C.~M. Stewart, A.~C. Walls, W.~W. Hannon,
  D.~Veesler, and J.~D. Bloom.
\newblock Deep mutational scans for ace2 binding, rbd expression, and antibody
  escape in the SARS-CoV-2 Omicron BA.1 and BA.2 receptor-binding domains.
\newblock {\em PLoS Pathogens}, 18(11):e1010951, 2022.

\bibitem{townsend2020representation}
J.~Townsend, C.~P. Micucci, J.~H. Hymel, V.~Maroulas, and K.~D. Vogiatzis.
\newblock Representation of molecular structures with persistent homology for
  machine learning applications in chemistry.
\newblock {\em Nature Communications}, 11(1):1--9, 2020.

\bibitem{walls2020structure}
A.~C. Walls, Y.-J. Park, M.~A. Tortorici, A.~Wall, A.~T. McGuire, and
  D.~Veesler.
\newblock Structure, function, and antigenicity of the SARS-CoV-2 spike
  glycoprotein.
\newblock {\em Cell}, 181(2):281--292, 2020.

\bibitem{wang2020topology}
M.~Wang, Z.~Cang, and G.-W. Wei.
\newblock A topology-based network tree for the prediction of protein--protein
  binding affinity changes following mutation.
\newblock {\em Nature Machine Intelligence}, 2(2):116--123, 2020.

\bibitem{wang2021vaccine}
R.~Wang, J.~Chen, K.~Gao, and G.-W. Wei.
\newblock Vaccine-escape and fast-growing mutations in the {United Kingdom, the
  United States, Singapore, Spain, India}, and other {COVID}-19-devastated
  countries.
\newblock {\em Genomics}, 113(4):2158--2170, 2021.

\bibitem{wang2021mechanisms}
R.~Wang, J.~Chen, and G.-W. Wei.
\newblock Mechanisms of SARS-CoV-2 evolution revealing vaccine-resistant
  mutations in Europe and America.
\newblock {\em The Journal of Physical Chemistry Letters}, 12:11850--11857,
  2021.

\bibitem{wang2020persistent}
R.~Wang, D.~D. Nguyen, and G.-W. Wei.
\newblock Persistent spectral graph.
\newblock {\em International Journal for Numerical Methods in Biomedical
  Engineering}, 36(9):e3376, 2020.

\bibitem{wang2023persistent}
R.~Wang and G.-W. Wei.
\newblock Persistent path Laplacian.
\newblock {\em Foundation of Data Science}, (5):26--55, 2023.

\bibitem{wang2021hermes}
R.~Wang, R.~Zhao, E.~Ribando-Gros, J.~Chen, Y.~Tong, and G.-W. Wei.
\newblock HERMES: Persistent spectral graph software.
\newblock {\em Foundations of Data Science}, 3(1):67--97, 2020.

\bibitem{wee2022persistent}
J.~Wee and K.~Xia.
\newblock Persistent spectral based ensemble learning (perspect-el) for
  protein--protein binding affinity prediction.
\newblock {\em Briefings in Bioinformatics}, 23(2), 2022.

\bibitem{wei2023topological}
G.-W. Wei.
\newblock Topological data analysis hearing the shapes of drums and bells.
\newblock {\em arXiv preprint arXiv:2301.05025}, 2023.

\bibitem{wei2021homotopy}
X.~Wei and G.-W. Wei.
\newblock Homotopy continuation for the spectra of persistent Laplacians.
\newblock {\em Foundations of Data Science}, 3(4):677, 2021.

\bibitem{wei2021persistent}
X.~Wei and G.-W. Wei.
\newblock Persistent sheaf Laplacians.
\newblock {\em arXiv preprint arXiv:2112.10906}, 2021.

\bibitem{KLXia:2014c}
K.~L. Xia and G.~W. Wei.
\newblock Persistent homology analysis of protein structure, flexibility and
  folding.
\newblock {\em International Journal for Numerical Methods in Biomedical
  Engineering}, 30:814--844, 2014.

\bibitem{xiang2001extending}
Z.~Xiang and B.~Honig.
\newblock Extending the accuracy limits of prediction for side-chain
  conformations.
\newblock {\em Journal of Molecular Biology}, 311(2):421--430, 2001.

\bibitem{xu2021conformational}
C.~Xu, Y.~Wang, C.~Liu, C.~Zhang, W.~Han, X.~Hong, Y.~Wang, Q.~Hong, S.~Wang,
  Q.~Zhao, et~al.
\newblock Conformational dynamics of sars-cov-2 trimeric spike glycoprotein in
  complex with receptor ACE2 revealed by cryo-EM.
\newblock {\em Science Advances}, 7(1):eabe5575, 2021.

\bibitem{zomorodian2005computing}
A.~Zomorodian and G.~Carlsson.
\newblock Computing persistent homology.
\newblock {\em Discrete \& Computational Geometry}, 33(2):249--274, 2005.

\bibitem{JustinCurry},
J.~Curry.
\newblock Sheaves, cosheaves and applications.
\newblock {The University of Pennsylvania}, 2014.

\end{thebibliography}

\clearpage 
\section{Appendix}

\begin{figure}[htbp]
    \centering
        \includegraphics[width=0.9\linewidth]{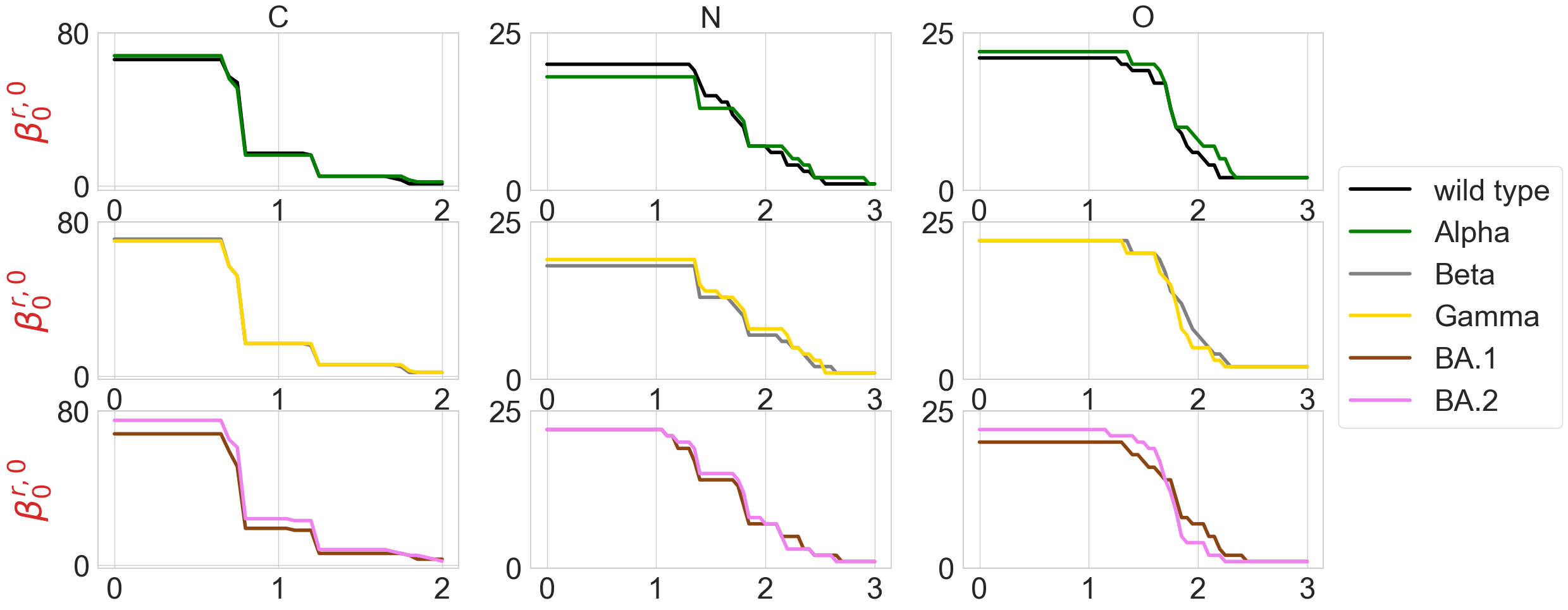}
    \caption{Illustration of persistent Betti numbers of element specific persistent Laplacian
    of the residue 501 mutation site at different filtration values, i.e., radii (unit: \AA).
			The wild type (PDB ID: 6M0J) and Alpha (PDB ID: 8DLK) are given in the first row. 
		The Beta (PDB ID: 8DLN) and Gamma (PDB ID: 8DLQ) are given in the second row. 
		The BA.1 (PDB ID: 7T9L) and BA.2 (PDB ID: 7XB0) are given the third row.
    }
    \label{mutation_es_beta_pl}
\end{figure}

\begin{figure}[htbp]
    \centering
        \includegraphics[width=0.9\linewidth]{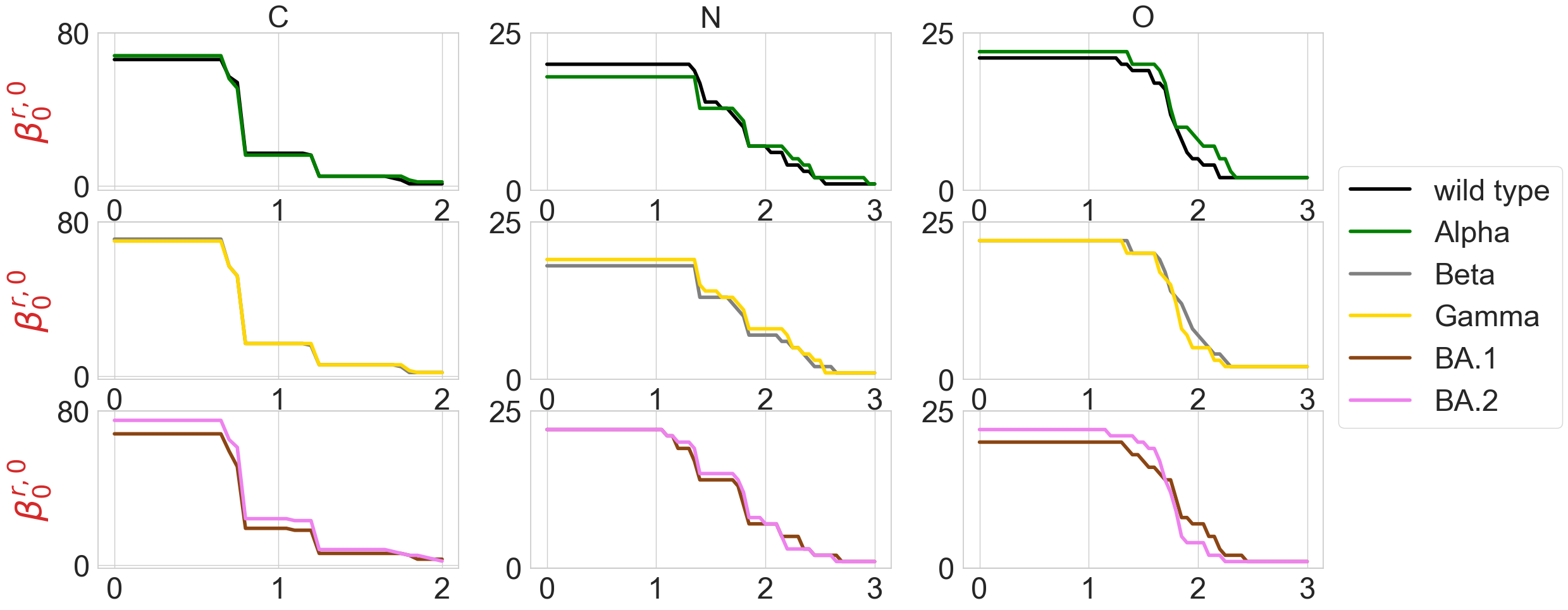}
    \caption{Illustration of persistent Betti numbers of element specific persistent sheaf Laplacian
    of the residue 501 mutation site at different filtration values, i.e., radii (unit: \AA). 		The wild type (PDB ID: 6M0J) and Alpha (PDB ID: 8DLK) are given in the first row. 
		The Beta (PDB ID: 8DLN) and Gamma (PDB ID: 8DLQ) are given in the second row. 
		The BA.1 (PDB ID: 7T9L) and BA.2 (PDB ID: 7XB0) are given the third row.
}
    \label{mutation_es_beta_psl}
\end{figure}

\begin{figure}[htbp]
    \centering
        \includegraphics[width=0.9\linewidth]{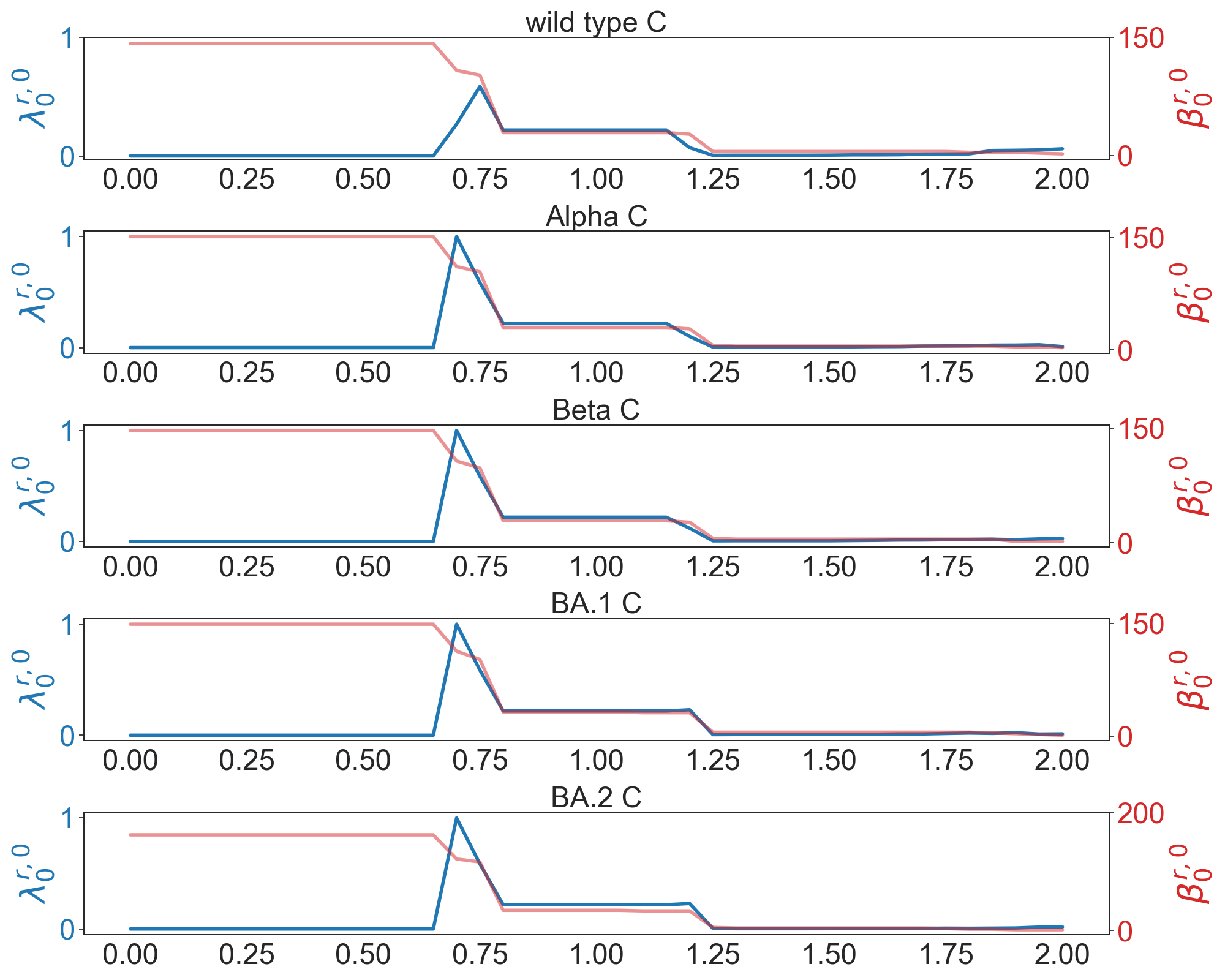}
    \caption{Illustration of persistent Betti numbers (red line) and the first nonzero eigenvalues (blue line) of persistent Laplacian
    of carbon atoms of the RBD binding site in RBD-ACE2 complex of the wild type (PDB ID: 6M0J), Alpha (PDB ID: 8DLK), Beta (PDB ID: 8DLN), BA.1 (PDB ID: 7T9L), and BA.2 (PDB ID: 7XB0) at different filtration values, i.e., radii (unit: \AA). 
    The graphs from top to bottom represent the results of the wild type, Alpha, Beta, BA.1, and BA.2 variants,  respectively.}
    \label{bindedsur}
\end{figure}

\begin{figure}[htbp]
    \centering
        \includegraphics[width=0.9\linewidth]{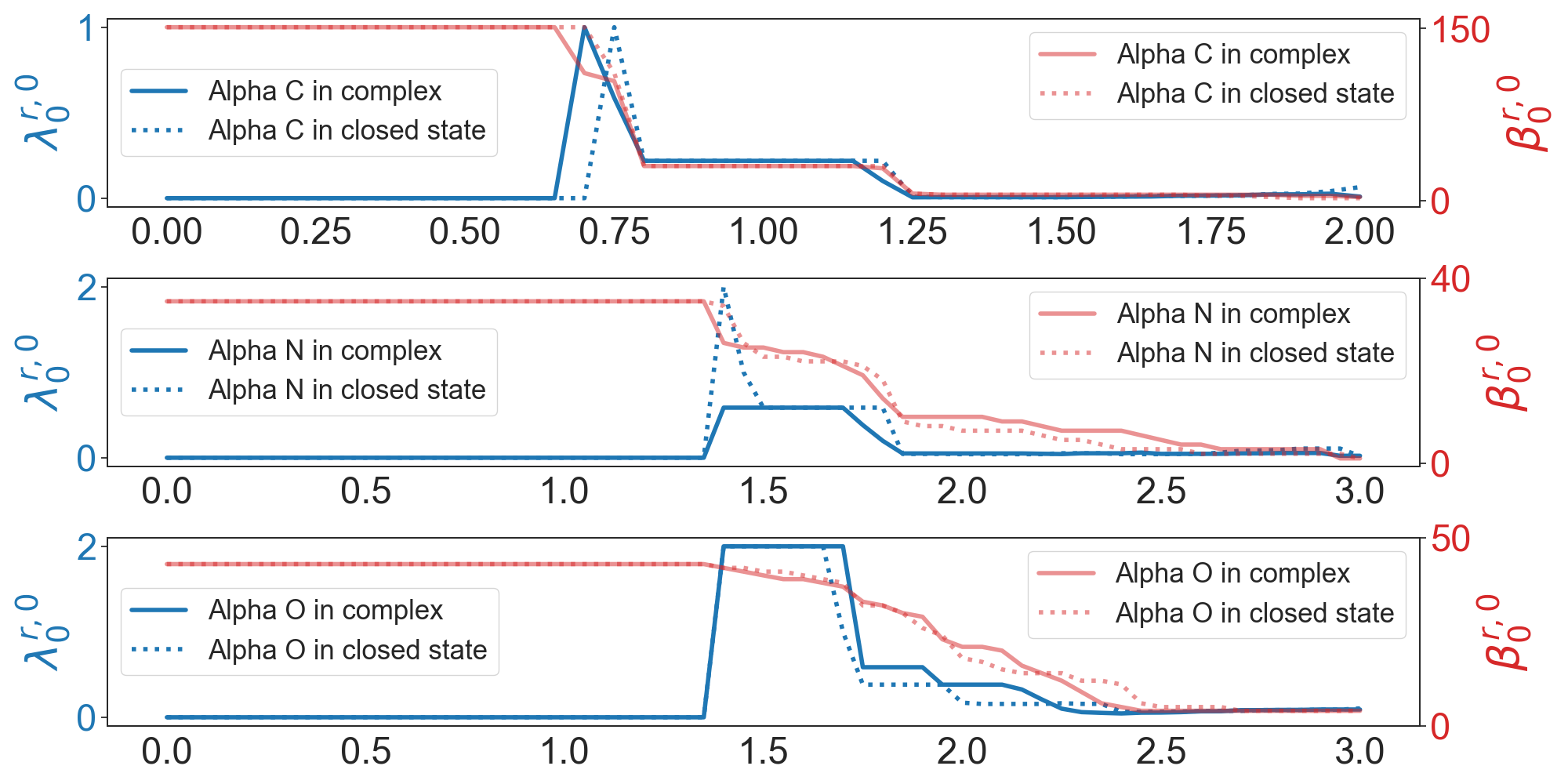}
    \caption{Illustration of persistent Betti numbers (red line) and the first nonzero eigenvalues (blue line) of persistent Laplacian
    of the RBD binding site of Alpha RBD-ACE2 complex (PDB ID: 8DLK) and closed state spike protein (PDB ID: 7LWS, Chain ID: A) at different filtration values, i.e., radii (unit: \AA). 
    The graphs from top to bottom represent the results of carbon atoms, nitrogen atoms, and oxygen atoms, respectively.}
    \label{bindedsurvsclosedsur_alpha}
\end{figure}

\begin{figure}[htbp]
    \centering
        \includegraphics[width=0.9\linewidth]{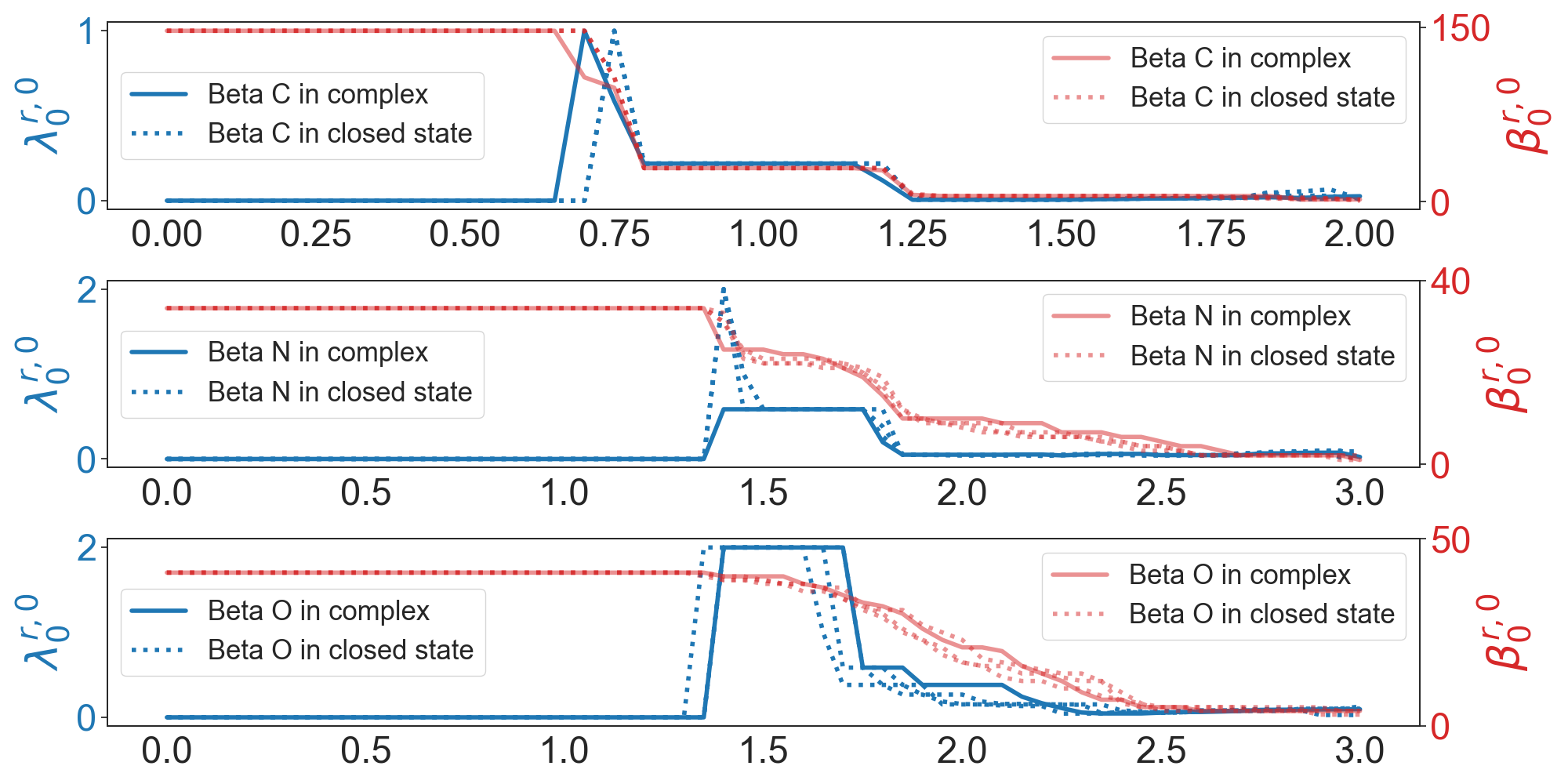}
    \caption{Illustration of persistent Betti numbers (red line) and the first  nonzero eigenvalues (blue line) of persistent Laplacian
    of the RBD binding site of Beta RBD-ACE2 complex (PDB ID: 8DLN) and closed state spike protein (PDB ID: 7LYM, Chain ID: A, B, C) at different filtration values, i.e., radii (unit: \AA). 
    The graphs from top to bottom represent the results of carbon atoms, nitrogen atoms, and oxygen atoms, respectively.}
    \label{bindedsurvsclosedsur_beta}
\end{figure}

\begin{figure}[htbp]
    \centering
        \includegraphics[width=0.9\linewidth]{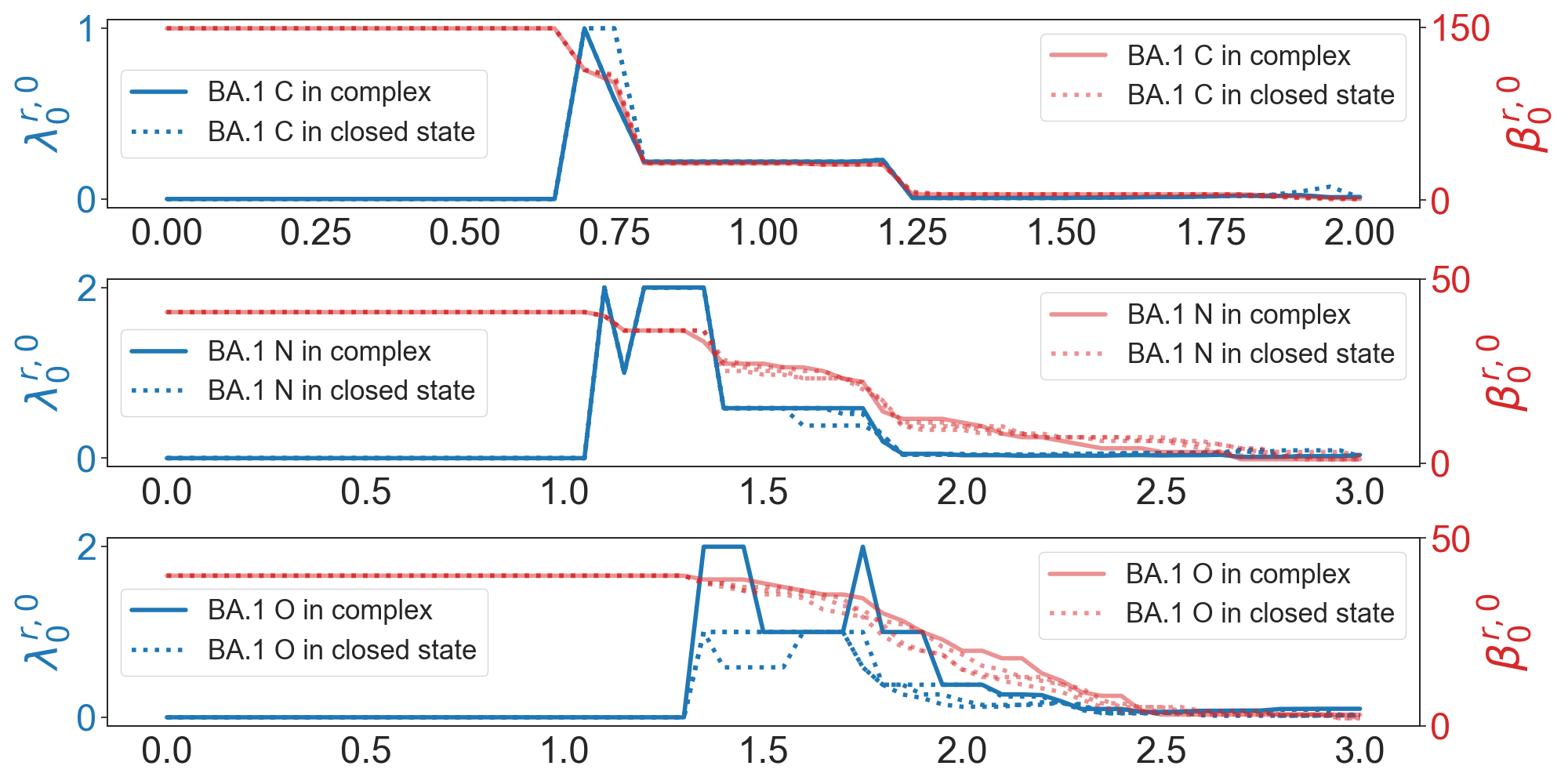}
    \caption{Illustration of persistent Betti numbers (red line) and the first nonzero eigenvalues (blue line) of persistent Laplacian
    of the RBD binding site of BA.1 RBD-ACE2 complex (PDB ID: 7T9L) and closed state spike protein (PDB ID: 7TF8, Chain ID: A, B, C) at different filtration values, i.e., radii (unit: \AA). 
    The graphs from top to bottom represent the results of carbon atoms, nitrogen atoms, and oxygen atoms, respectively.}
    \label{bindedsurvsclosedsur_ba1}
\end{figure}

\begin{figure}[htbp]
    \centering
        \includegraphics[width=0.9\linewidth]{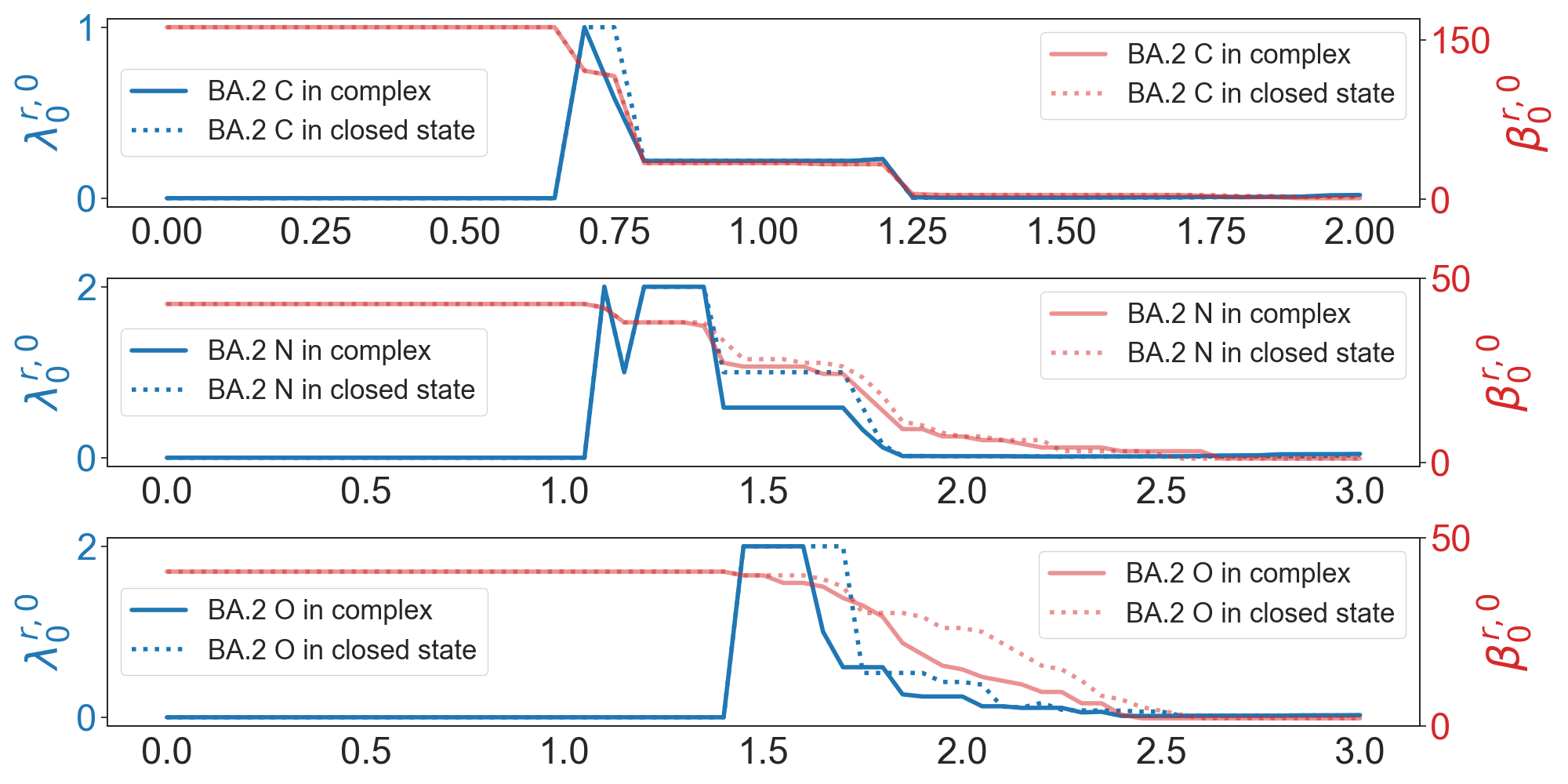}
    \caption{Illustration of persistent Betti numbers (red line) and the first  nonzero eigenvalues (blue line) of persistent Laplacian
    of the RBD binding site of BA.2 RBD-ACE2 complex (PDB ID: 7XB0) and closed state spike protein (PDB ID: 7XIX, Chain ID: A) at different filtration values, i.e., radii (unit: \AA). 
    The graphs from top to bottom represent the results of carbon atoms, nitrogen atoms, and oxygen atoms, respectively.}
    \label{bindedsurvsclosedsur_ba2}
\end{figure}

\begin{figure}[htbp]
    \centering
        \includegraphics[width=0.9\linewidth]{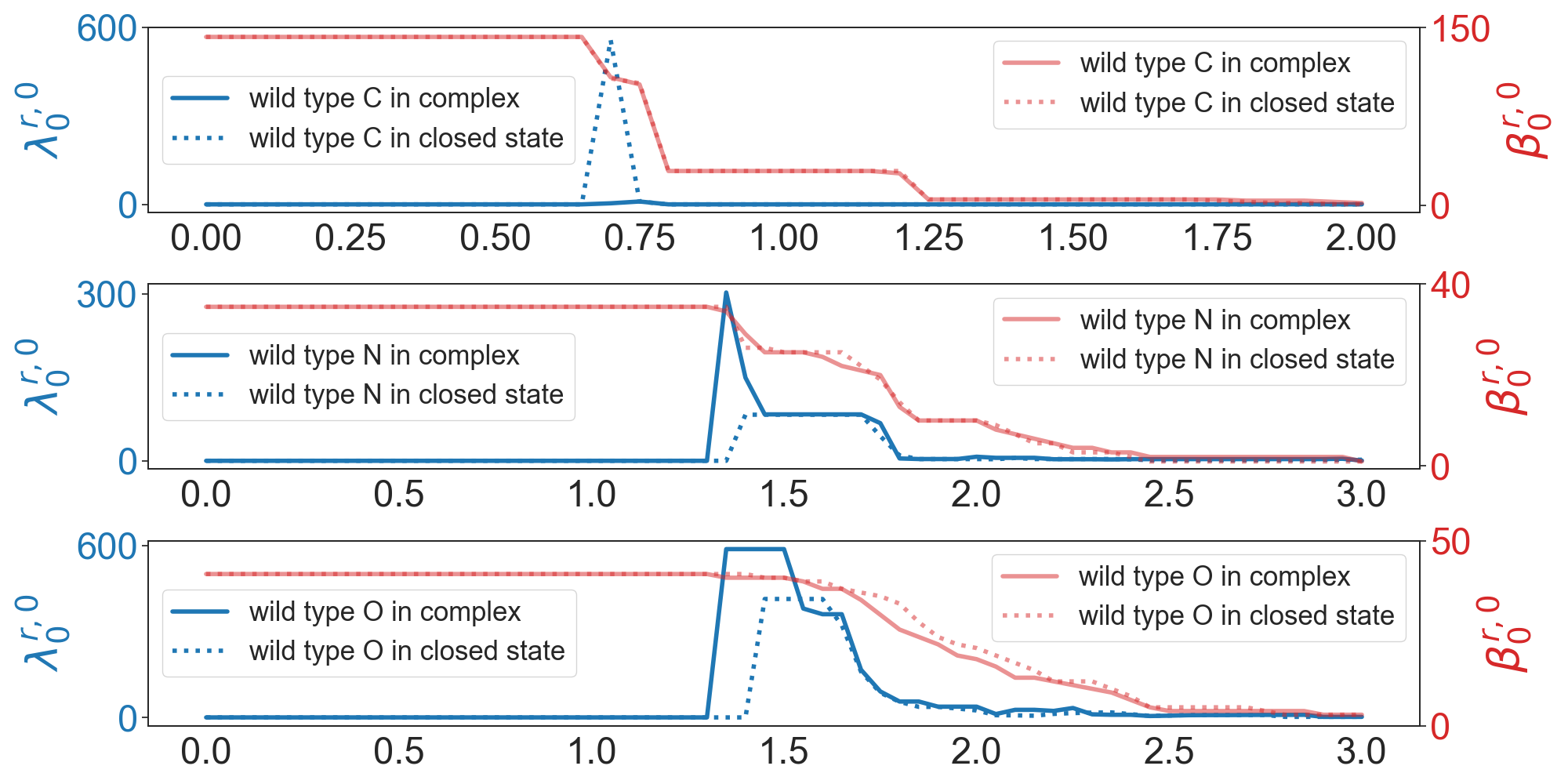}
    \caption{Illustration of persistent sheaf Betti numbers (red line) and the first  nonzero eigenvalues (blue line) of persistent sheaf Laplacian
    of the RBD binding site of the wild type RBD-ACE2 complex (PDB ID: 6M0J) and closed state spike protein (PDB ID: 7DF3, Chain ID: A) at different filtration values, i.e., radii (unit: \AA). 
    The graphs from top to bottom represent the results of carbon atoms, nitrogen atoms, and oxygen atoms respectively.}
    \label{bindedsurvsclosedsur_wild_psl}
\end{figure}

\begin{figure}[htbp]
    \centering
        \includegraphics[width=0.9\linewidth]{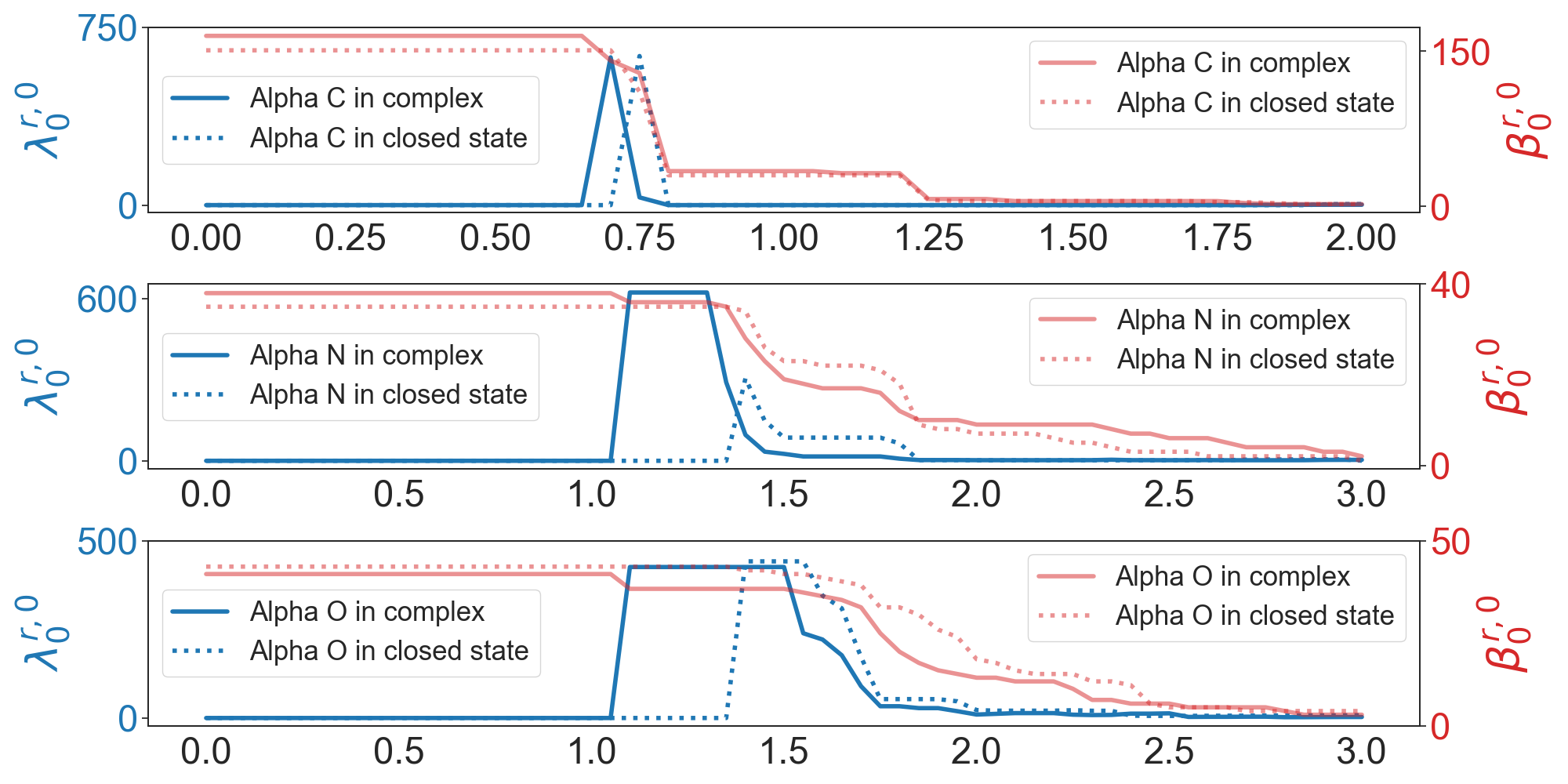}
    \caption{Illustration of persistent sheaf Betti numbers (red line) and the first  nonzero eigenvalues (blue line) of persistent sheaf Laplacian
    of the RBD binding site of Alpha RBD-ACE2 complex (PDB ID: 8DLK) and closed state spike protein (PDB ID: 7LWS, Chain ID: A) at different filtration values, i.e., radii (unit: \AA). 
    The graphs from top to bottom represent the results of carbon atoms, nitrogen atoms, and oxygen atoms, respectively.}
    \label{bindedsurvsclosedsur_alpha_psl}
\end{figure}

\begin{figure}[htbp]
    \centering
        \includegraphics[width=0.9\linewidth]{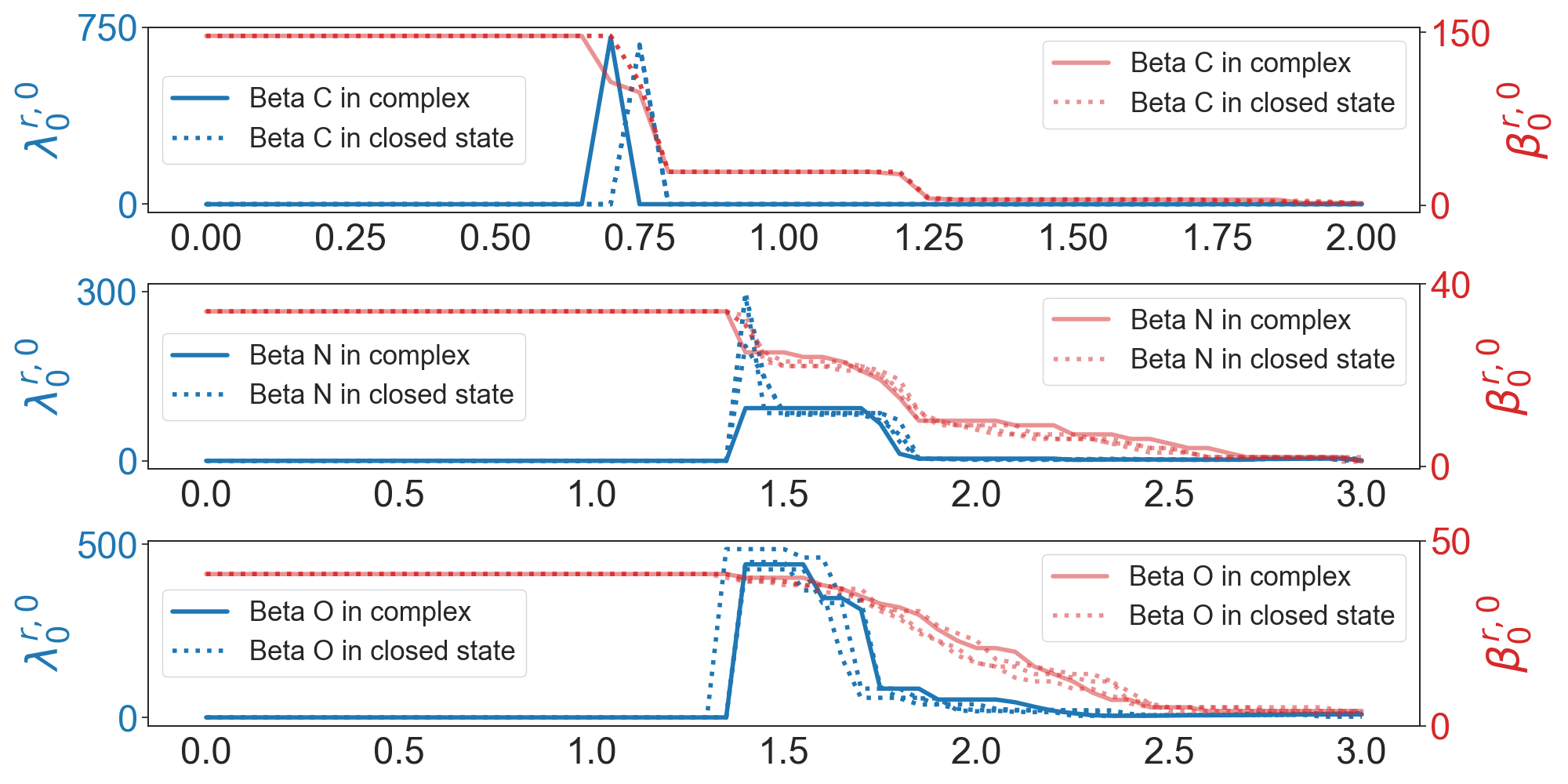}
    \caption{Illustration of persistent sheaf Betti numbers (red line) and the first  nonzero eigenvalues (blue line) of persistent sheaf Laplacian
    of the RBD binding site of Beta RBD-ACE2 complex (PDB ID: 8DLN) and closed state spike protein (PDB ID: 7LYM, Chain ID: A, B, C) at different filtration values, i.e., radii (unit: \AA). 
    The graphs from top to bottom represent the results of carbon atoms, nitrogen atoms, and oxygen atoms, respectively.}
    \label{bindedsurvsclosedsur_beta_psl}
\end{figure}

\begin{figure}[htbp]
    \centering
        \includegraphics[width=0.9\linewidth]{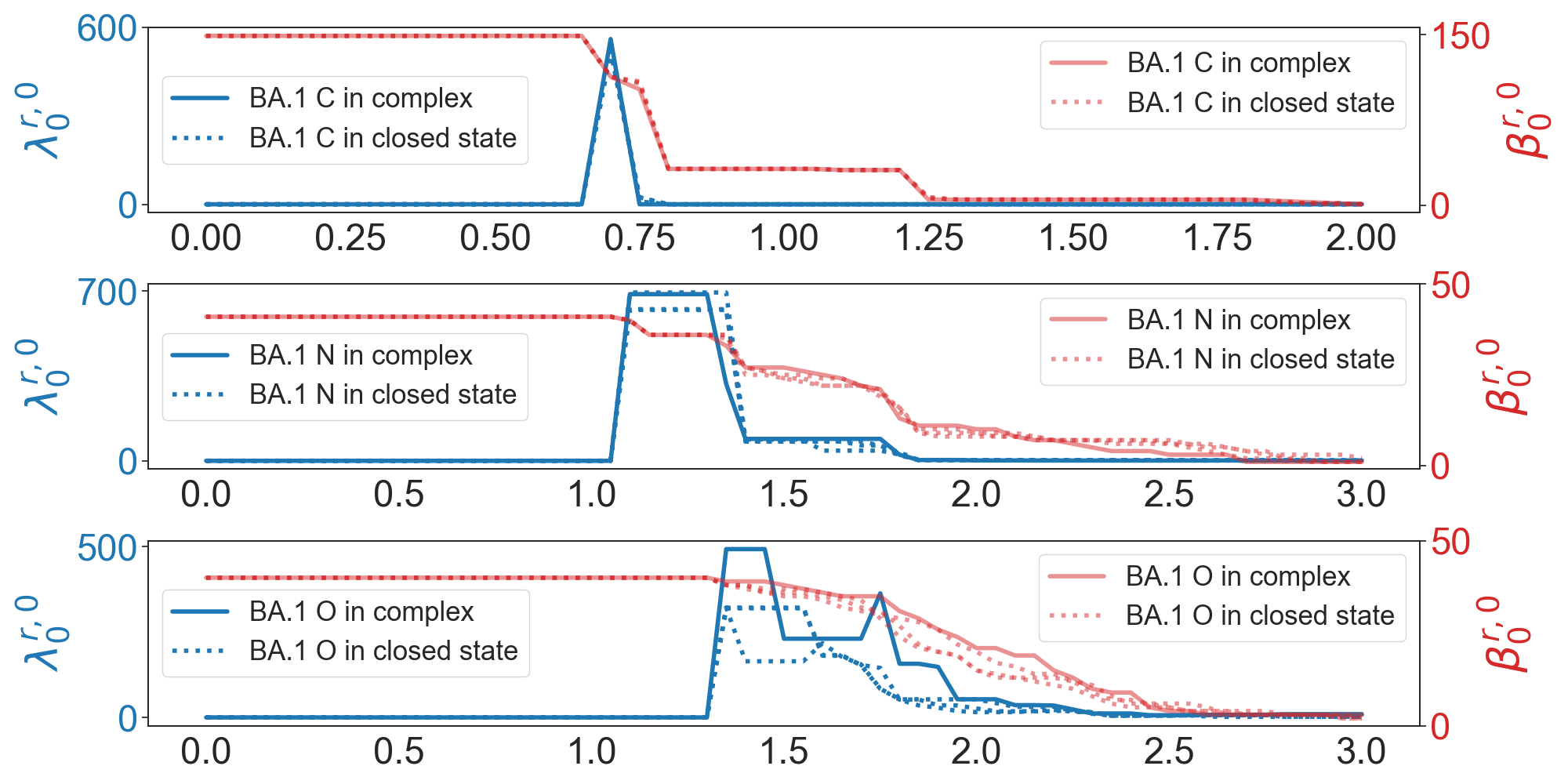}
    \caption{Illustration of persistent sheaf Betti numbers (red line) and the first nonzero eigenvalues (blue line) of persistent sheaf Laplacian
    of the RBD binding site of BA.1 RBD-ACE2 complex (PDB ID: 7T9L) and closed state spike protein (PDB ID: 7TF8, Chain ID: A, B, C) at different filtration values, i.e., radii (unit: \AA). 
    The graphs from top to bottom represent the results of carbon atoms, nitrogen atoms, and oxygen atoms, respectively.}
    \label{bindedsurvsclosedsur_ba1_psl}
\end{figure}

\begin{figure}[htbp]
    \centering
    \begin{subfigure}[b]{0.9\textwidth}
        \centering
        \includegraphics[width=0.9\linewidth]{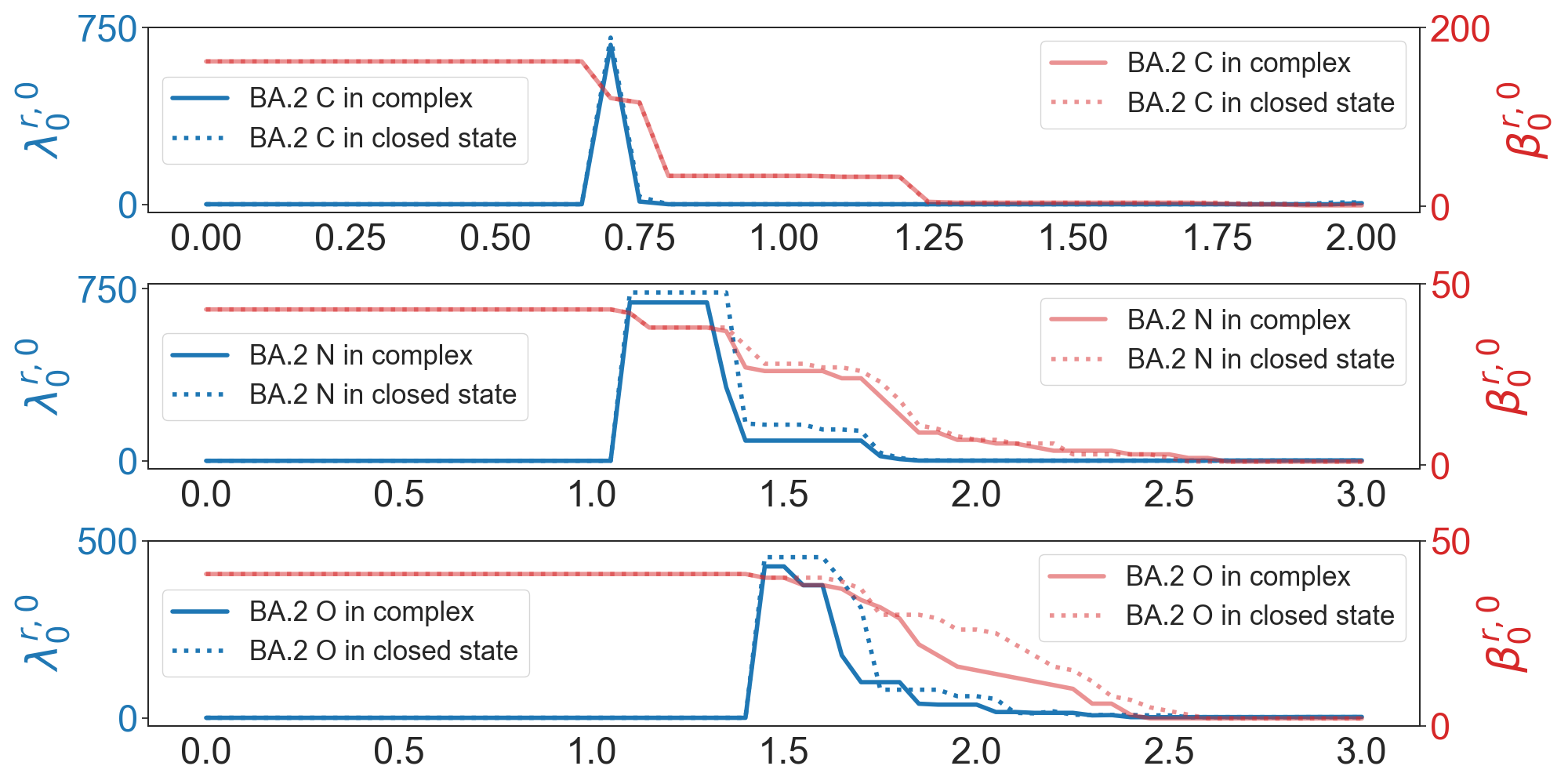}
    \end{subfigure}
    \caption{Illustration of persistent sheaf Betti numbers (red line) and the first  nonzero eigenvalues (blue line) of persistent sheaf Laplacian
    of the RBD binding site of BA.2 RBD-ACE2 complex (PDB ID: 7XB0) and closed state spike protein (PDB ID: 7XIX, Chain ID: A) at different filtration values, i.e., radii (unit: \AA). 
    The graphs from top to bottom represent the results of carbon atoms, nitrogen atoms, and oxygen atoms, respectively.}
    \label{bindedsurvsclosedsur_ba2_psl}
\end{figure}

In the appendix, we provide  the information of PDB structures we used   and
additional topological analysis using persistent Laplacian and persistent sheaf Laplacian.  
 The information of PDB structures is given in Table \ref{pdbinfo}.
PTL results are given in Figures 
\ref{mutation_es_beta_pl},
\ref{mutation_es_beta_psl}, 
\ref{bindedsur}, 
\ref{bindedsurvsclosedsur_alpha}, 
\ref{bindedsurvsclosedsur_beta}, 
\ref{bindedsurvsclosedsur_ba1}, 
\ref{bindedsurvsclosedsur_ba2}, 
\ref{bindedsurvsclosedsur_wild_psl}, 
\ref{bindedsurvsclosedsur_alpha_psl}, 
\ref{bindedsurvsclosedsur_beta_psl}, 
\ref{bindedsurvsclosedsur_ba1_psl}, and 
\ref{bindedsurvsclosedsur_ba2_psl}. 

Specifically, Figures \ref{mutation_es_beta_pl}, and \ref{mutation_es_beta_psl} are element-specific analysis of the wide type, Alpha, Beta, Gamma, BA.1, and BA.2 using LP and PSL, respectively.   
Figure \ref{bindedsur} presents carbon specific analysis of the wide type, Alpha, Beta, Gamma, BA.1, and BA.2.   
Figures \ref{bindedsurvsclosedsur_alpha}, 
\ref{bindedsurvsclosedsur_beta}, 
\ref{bindedsurvsclosedsur_ba1}, and 
\ref{bindedsurvsclosedsur_ba2}  demonstrate the PL analysis of Alpha, Beta, BA.2, and BA.2, respectively. The spectral analysis of three major types of elements, namely carbon atoms, nitrogen atoms, and oxygen atoms, is presented in these figures. 
Finally, 
\ref{bindedsurvsclosedsur_wild_psl}, 
\ref{bindedsurvsclosedsur_alpha_psl}, 
\ref{bindedsurvsclosedsur_beta_psl}, 
\ref{bindedsurvsclosedsur_ba1_psl}, and 
\ref{bindedsurvsclosedsur_ba2_psl} illustrate the PSL analysis of the wide type, Alpha, Beta, BA.2, and BA.2, respectively. These figures display the spectral analysis of three major types of elements, namely carbon atoms, nitrogen atoms, and oxygen atoms.

\begin{center}
    \begin{tabular}{ |c|c|c|c| } 
     \hline
     PDB ID & Method & Resolution (unit: \AA) & Description \\ 
     \hline
    6M0J\cite{lan2020structure}  & X-ray diffraction & 2.45 & wild type RBD-ACE2 \\ 
     \hline
    8DLK\cite{mannar2022sars}  & Electron microscopy & 3.04 & Alpha RBD-ACE2 \\ 
     \hline
     8DLN\cite{mannar2022sars} & Electron microscopy & 3.04 & Beta RBD-ACE2 \\ 
     \hline 
    8DLQ\cite{mannar2022sars} & Electron microscopy & 2.77 & Gamma RBD-ACE2\\
    \hline 
    7T9L\cite{mannar2022omicron} & Electron microscopy & 2.66 & BA.1 RBD-ACE2\\
    \hline 
    7XB0\cite{li2022structural} & X-ray diffraction & 2.90 & BA.2 RBD-ACE2\\ 
    \hline 
    7DF3\cite{xu2021conformational} & Electron microscopy & 2.70 & wild type spike\\
    \hline 
    7LWS\cite{gobeil2021effect} &Electron microscopy & 3.22 & Alpha spike\\
    \hline 
    7LYM\cite{gobeil2021effect} & Electron microscopy & 3.57 & Beta spike\\
    \hline 
    7TF8\cite{gobeil2022structural} & Electron microscopy & 3.36 & BA.1 spike\\ 
    \hline 
    7XIX\cite{cao2022ba} & Electron microscopy & 3.25 & BA.2 spike\\
    \hline
    \end{tabular}
    \captionof{table}{Information of PDB 3D structures used in this work.}
    \label{pdbinfo}
\end{center}

\end{document}